\documentclass[aps,prd,floatfix,amsmath,amssymb,preprint,eqsecnum,nofootinbib]{revtex4}
\usepackage{graphicx}
\usepackage{dcolumn}
\usepackage{bm}
\usepackage{color}

\newcommand{\beq}{\begin{equation}}
\newcommand{\eeq}{\end{equation}}
\newcommand{\ba}{\begin{array}{ccc}}
\newcommand{\ea}{\end{array}}
\newcommand{\nn}{\nonumber \\}

\def\bea{\begin{eqnarray}}
\def\eea{\end{eqnarray}}

\usepackage{setspace} 
\usepackage{amsmath}
\usepackage{amssymb}
\begin{document}

\title{What can gauge-gravity duality teach us\\ about condensed matter physics?}

\author{Subir Sachdev}
\affiliation{Department of Physics, Harvard University, Cambridge MA
02138}

\date{\today \\
\vspace{1in}}

\begin{abstract}
I discuss the impact of gauge-gravity duality on our understanding of two classes of systems: 
{\em conformal\/} quantum matter and {\em compressible\/} quantum matter. 

The first conformal
class includes systems, such as the boson Hubbard model in two spatial dimensions, which display quantum critical
points described by conformal field theories. Questions associated with non-zero temperature dynamics and transport
are difficult to answer using conventional field theoretic methods. I argue that many of these can be addressed
systematically using gauge-gravity duality, and discuss the prospects for reliable computation of low frequency correlations.

Compressible quantum matter is characterized by the smooth dependence of the charge density, associated with a global U(1)
symmetry, upon a chemical potential. Familiar examples are solids, superfluids, and Fermi liquids, but there are more exotic possibilities
involving deconfined phases of gauge fields in the presence of Fermi surfaces.
I survey the compressible systems studied using gauge-gravity duality, and discuss their relationship
to the condensed matter classification of such states. The gravity methods offer hope of a deeper understanding of exotic
and strongly-coupled compressible quantum states. 
\end{abstract}

\maketitle

\section{Introduction}
\label{sec:intro}

One of the remarkable developments to emerge from research in string theory in the past decade is the idea of gauge-gravity 
duality \cite{MAGOO}.
This is an equivalence between a quantum field theory in $D$ spacetime dimensions, and a quantum theory of gravity
in $D+1$ spacetime dimensions. The $D$-dimensional theory does not have a gravitational force, and is to be viewed
as a `hologram' of the $D+1$ dimensional theory. This is a true equivalence between theories, and not a projection to
a restricted portion of Hilbert space: the number of degrees of freedom of the $D$-dimensional theory equal those
of the $(D+1)$-dimensional theory with gravity. It is thus in the spirit of early ideas of Bekenstein and 
others \cite{beken,hawking,susskind,thooft} that
in theories of quantum gravity, the entropy of a 
region of spacetime is proportional to its spatial surface area (and not its spatial volume).

A powerful feature of the duality is that it is also a strong-weak coupling duality: it maps a strongly-coupled theory in
$D$ dimensions to a weakly-coupled theory in $D+1$ dimensions, and vice versa. It has therefore, justifiably, raised hopes
for understanding new classes of strongly coupled quantum field theories in $D$ dimensions. This hope extends
to researchers, like condensed matter physicists, who had no previous interest in theories of quantum gravity.

The initial examples \cite{MAGOO} 
of gauge-gravity duality appeared in highly supersymmetric gauge theories {\em e.g.\/} the SU($N$) Yang-Mills gauge
theory with maximal supersymmetry ($\mathcal{N}=4$) in $D=4$. This is dual to a string theory on a $5$-dimensional
space with constant negative curvature: anti-de Sitter space, abbreviated AdS$_5$. In the low energy limit, the string theory reduces
to a gravity theory on AdS$_5$ also with maximal supersymmetry. Clearly, neither of these models are of any
specific interest to condensed matter physics. However, since then it has become clear that this is but the simplest of a general
class of duality between conformally invariant field theories (CFTs) in $D$ dimensions and theories of gravity in AdS$_{D+1}$.
Strong evidence for such a duality has also appeared for CFTs without any supersymmetry \cite{yin1,yin2}. As we will review in Section~\ref{sec:cond1},
CFTs appear at quantum phase transitions of generic models relevant for experiments in condensed matter and ultracold
atomic gases, and some of these are directly in the classes for which gauge-gravity duality has been studied.

It is perhaps useful here to make an analogy with earlier developments in the solution of strongly interacting many body systems.
Starting with the work of Bethe in 1931 \cite{bethe}, a wide class of solvable quantum many body systems were discovered in one spatial dimension
($D=2$). These models are `integrable' in that they have an infinite number of conservation laws, and explicit wavefunctions can be
written down for all eigenstates using the Bethe ansatz. However, integrability invariably required fine-tuning of the structure
of the Hamiltonian, and no one expected that integrable systems would be directly applicable to any experiments.
Nevertheless, the structure of the integrable models and their excitations taught us a great deal about quantum many body
physics in $D=2$, and these insights were crucial in the development of the far more general effective
field theory of the `Tomonaga-Luttinger'
liquid \cite{haldane}. The latter theory applies to generic interacting systems in $D=2$, and has found
 numerous experimental applications. 

The hope is that a similar success will eventually
be achieved using gauge-gravity duality, for strongly interacting quantum systems in $D>2$. We have the analog
of the Bethe ansatz: a rapidly-increasing class of special models with known gravity duals. Much 
research \cite{gubsertasi,hartnoll1,hartnoll2,herzog,horowitz,mcgreevy,tasi,statphys,greece} is now directed
towards generalizing the insights obtained from these solvable models to a general effective field theory approach 
to strongly interacting quantum systems using holography.

I will review applications of gauge-gravity duality to two broad classes of condensed matter systems, which I 
denote ``conformal quantum matter'' and ``compressible quantum matter''. 

In the first ``conformal'' class I include models
with quantum critical points described by CFTs. Here the relevance of the holographic approach is evident, as
the equivalence to dual gravity models has been explicitly demonstrated.
Much has been learnt about such CFTs using traditional field-theoretic methods (such as the $\epsilon$ expansion
of Wilson and Fisher \cite{wf}). However, there are key questions about their long-time correlations at non-zero temperatures ($T$)
which cannot be described in a controlled manner by these methods. Numerical studies also fail for such questions because
of difficulties associated with the ``sign'' problem: quantum simulations require summing over intermediate
states whose weights are not positive real numbers. Remarkably, holography does yield useful results for such long-time
correlations, including the specific numerical values of transport co-efficients and damping rates: these results will be
discussed in Section~\ref{sec:cft}.

The second class of ``compressible quantum matter'' will be defined more precisely in Section~\ref{sec:nfl}.
But as the name indicates, these are systems which have a non-zero and finite compressibility at $T=0$. Familiar examples
of compressible states are Fermi liquids, solids, and superfluids. However, much modern condensed matter
research has focused on the search for new types of compressible states, motivated primarily by the ``strange metal''
behavior of numerous correlated electron systems \cite{physicstoday}. Current studies using holography have produced
many examples of compressible quantum matter, although their place in the traditional condensed matter classification
has not yet been fully established. These issues are the focus of active research, as I will describe in Section~\ref{sec:nfl}.

I close this section by noting that the above classification implicitly assumes $D>2$. The case $D=2$ is special:
Tomonaga-Luttinger liquids are {\em both\/} conformal and compressible, and this is partly responsible for their simplicity.
I will implicitly assume $D>2$ in the remainder of the paper, where the two classes are distinct and far more complicated,
and have properties quite different from Tomonaga-Luttinger liquids.

\section{Conformal quantum matter}
\label{sec:cft}

\subsection{Condensed matter analysis}
\label{sec:cond1}

\begin{figure}[htbp]
  \centering
  \includegraphics[width=2.in]{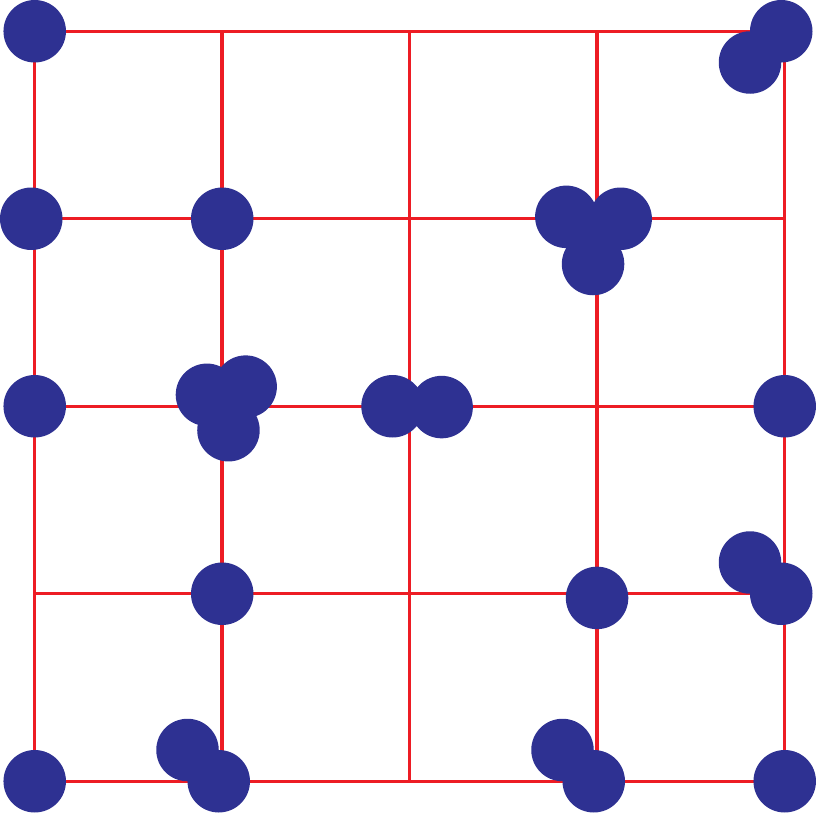} \quad\quad\quad
  \includegraphics[width=2in]{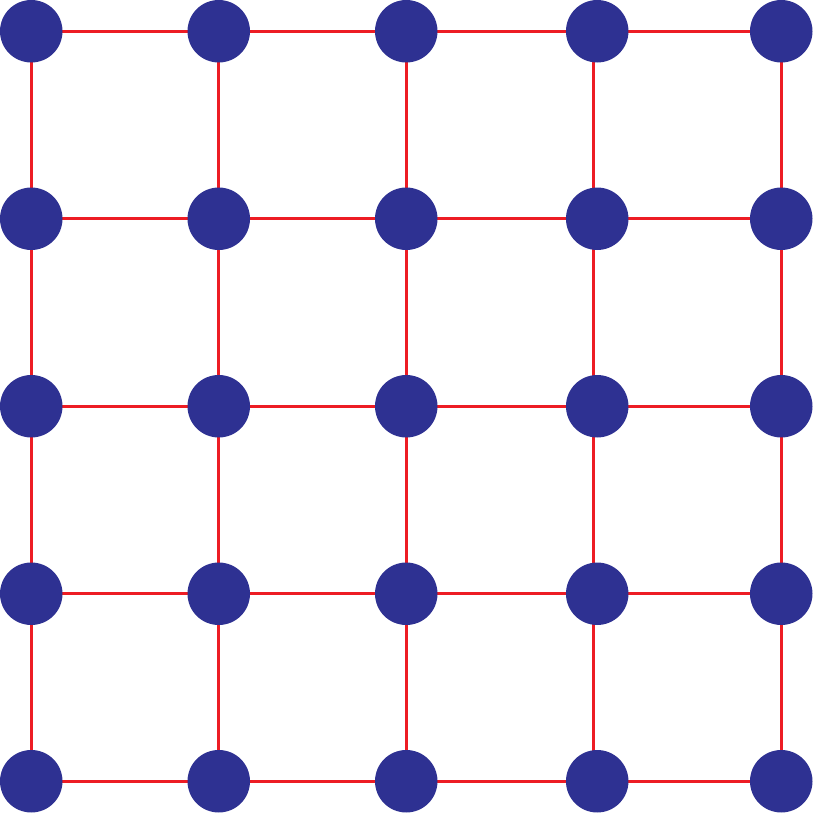}
  \caption{Superfluid (left) and insulating states of bosons with repulsive interactions on a square
  lattice at $f=1$.}
  \label{integer}
\end{figure}

Let us begin by describing the simplest model which realizes a CFT, realizable in a laboratory \cite{bloch,chin}. 
Consider spinless bosons
hopping on a lattice of sites, $i$, with short-range repulsive interactions. With a boson annihilation operator $b_i$, such
physics is usefully captured by the Hubbard Hamiltonian
\beq
H_b = - w \sum_{\langle ij \rangle} \left( b^\dagger_i b_j + b_j^\dagger b_i \right) + \frac{U}{2} \sum_i n_i (n_i - 1)
\label{hubbard}
\eeq
where $b_i$ is the canonical boson annihilation operator, 
$n_i = b^\dagger_i b_i$ is the boson number operator, $w$ is the hopping matrix element between nearest-neighbor
sites, and $U$ is the on-site repulsive energy between a pair of bosons. Let us assume that the average boson density is exactly 1
per site, and examine the ground state as a function of the dimensionless parameter $U/w$. For large $U/w$, the boson repulsion dominates
and so the bosons avoid each other by localizing on separate sites. This leads to the insulating state shown in Fig.~\ref{integer}: any motion of bosons requires placing at least two on a site, and this is suppressed by the repulsive energy. In the opposite limit of small $U/w$, we can treat
the system as a nearly free Bose gas, and this undergoes Bose-Einstein condensation to a the superfluid state also illustrated in Fig.~\ref{integer}.
In this state, a snapshot of the wavefunction shows large number fluctuations on each lattice site, and so a supercurrent is able to flow without dissipation. 

The distinct ground states in the limit of small and large $U/w$ implies they cannot be smoothly connected: they are separated by
a quantum phase transition at an intermediate critical point. More generally we denote parameters like $U/w$ by a generic coupling $g$,
and the quantum critical point appears at $g=g_c$. The nature of the ground state near and at $g=g_c$ is well understood,
and a pedagogical description appears in the Supplementary Material.
For our purposes, the
most important property is that the ground state and its low energy excitations are efficiently described by a universal quantum field 
theory \cite{fwgf}.
By `universal' we mean that the same field theory applies for a large class of models of the superfluid-insulator transition, and not just
the Hubbard model in Eq.~(\ref{hubbard}). The field theory is expressed in terms of a complex field $\psi (r,t)$, which is just the continuum
limit of the boson annihilation operator $b_i$, with $r$ the $D-1$ dimensional spatial co-ordinate, and $t$ the time. 
The action for the field theory is (see Supplementary Material)
\beq
\mathcal{S}_b = \int d^{D-1}r\, dt \left( - |\partial_t \psi|^2 + v^2 |\nabla_r \psi|^2 + (g-g_c) |\psi|^2 + u |\psi|^4 \right); \label{sb}
\eeq
note that its couplings change as a function of the coupling $g$. This low energy quantum field theory already has an `emergent' 
symmetry not shared by the Hubbard model: it is invariant under Lorentz transformations, with the velocity $v$ playing the role of the velocity
of light. Here $v$ is a sound velocity, and its value is determined by $w$, $U$, and the lattice spacing. 

The symmetry of the theory
becomes much larger at the quantum critical point at $g=g_c$. Here, as may be familiar to some readers from the theory of 
second order phase transitions, the theory is {\em scale invariant\/}. Specifically, the structure of the quantum correlations remain
invariant under the rescaling transformation
\beq
t \rightarrow t/b \quad,\quad r \rightarrow r/b \label{scal}
\eeq
where $b$ is the rescaling factor. Formally, this scale invariance arises because the quantum critical point is realized as a fixed point of the
renormalization group transformation. Actually, we are not done with the set of emergent symmetries. Quantum field theories which
are Lorentz invariant and which obey certain `hyperscaling' properties, 
are also invariant under {\em conformal\/} transformations of spacetime. These are transformations
which preserve the Lorentzian metric of spacetime upto an overall factor, and so preserve all angles. Specifically, the
Lorentzian metric rescales under conformal transformations
\beq
(-v^2 dt^2 + dr^2) \rightarrow (-v^2 dt^2 + dr^2)/ b^2 (r,t)  \label{conform}
\eeq
where the rescaling factor $b$ can be a certain set of smooth functions of spacetime; the rescaling transformation is a special
case of a conformal transformation. The critical point of $\mathcal{S}_b$, describing the superfluid-insulator
quantum phase transition, is invariant under such conformal transformations, and is so a CFT. We will restrict further attention to the case of 2 spatial
dimensions, with $D=3$, when this CFT is strongly coupled. The CFTs for $D\geq 4$ are essentially free field theories, and we don't need sophisticated methods to analyze them.

As we noted earlier, many properties of the strongly coupled CFT at $T=0$ are well understood. The main tool is the renormalization group,
and its realization in the context of various analytic and numerical expansions---these are well-established methods which I will not dwell on here.
However, all of these methods fail for certain key questions on the dynamics at $T>0$.  

\begin{figure}[htbp]
  \centering
  \includegraphics[width=4.5in]{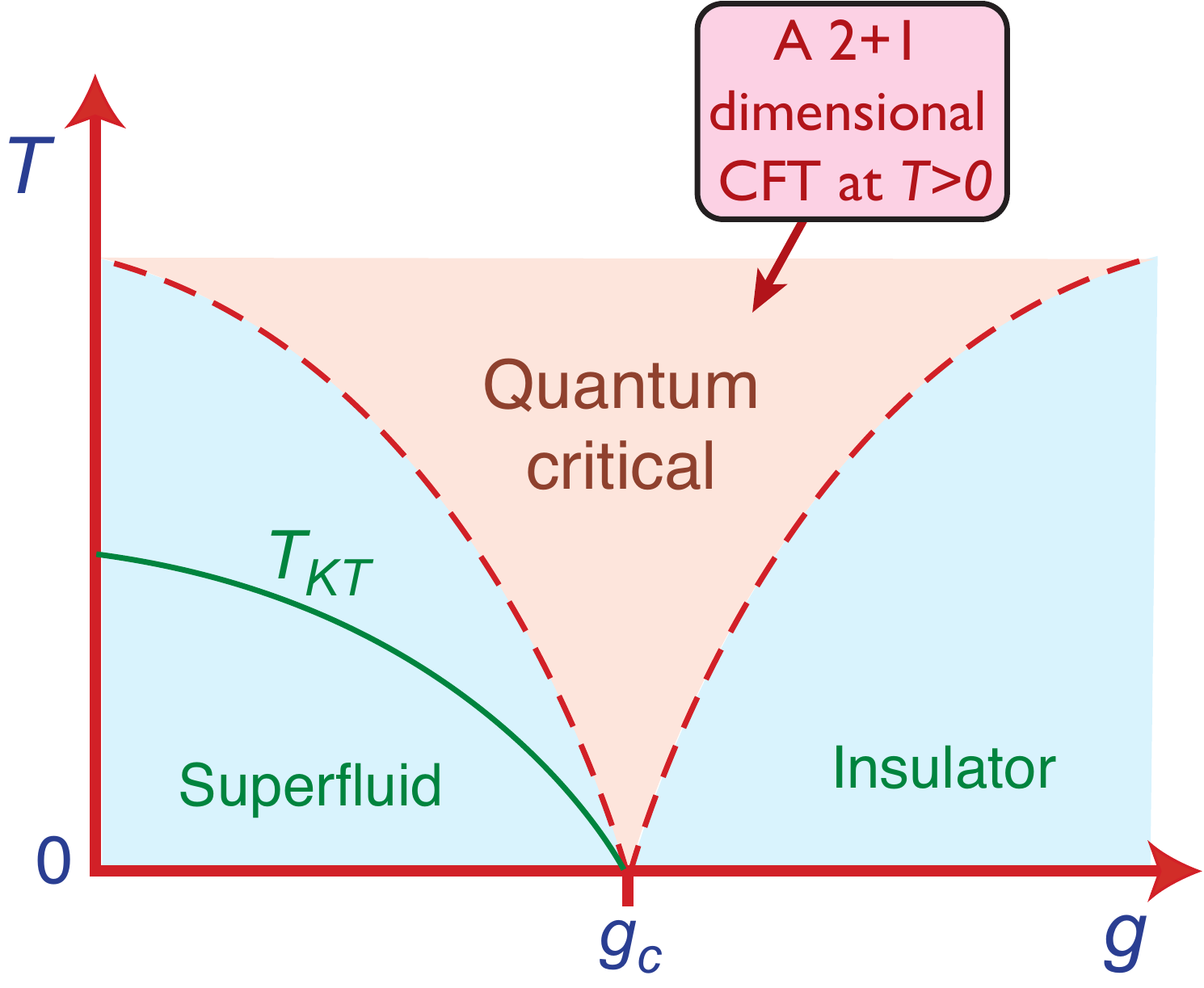}
  \caption{Phase diagram of the superfluid-insulator transition in two spatial dimensions ($D=3$). The quantum critical point is
  at $g=g_c$, $T=0$. The dashed lines are crossovers,
  while the full line is a phase transition at the Kosterlitz-Thouless temperature $T_{KT}>0$.}
  \label{qcfig}
\end{figure}

\begin{figure}[htbp]
  \centering
  \includegraphics[width=4.5in]{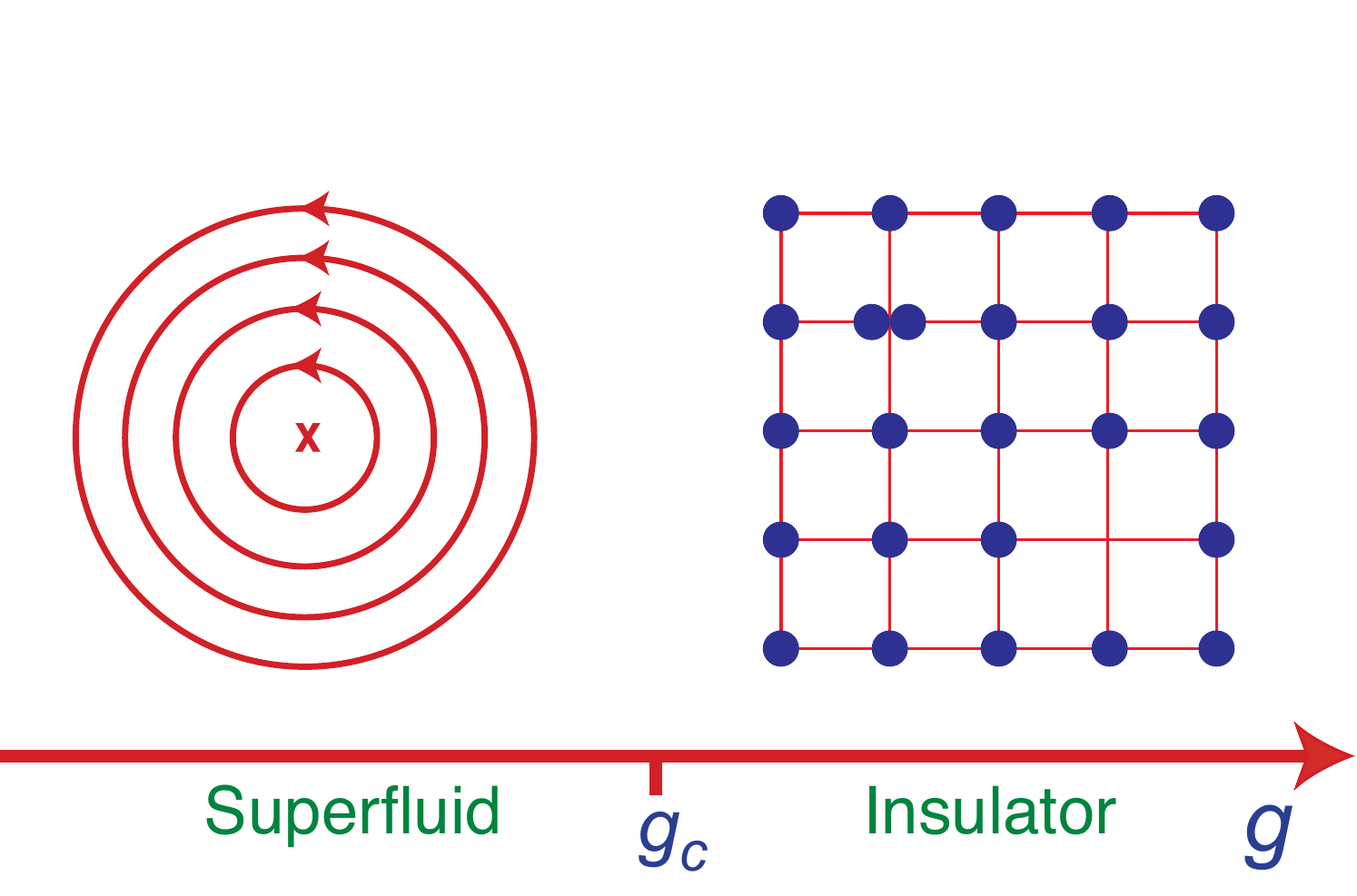}
  \caption{Excitations of the superfluid and the insulator. The excitations of the superfluid 
  are vortices and anti-vortices, which can be viewed
  as point-like particles centered at the X. A single particle and single hole excitations of the insulator are shown on the right.}
  \label{excite}
\end{figure}

To describe these questions, we need the phase
diagram of the model at $T>0$ \cite{book}, shown in Fig.~\ref{qcfig}.
There is much interesting physics associated with the many distinct features of this phase diagram, but for now the reader
is asked to focus on the distinction between the blue- and pink-shaded regions. In the blue-shaded regions, the physics can be
described in terms of the familiar excitations of either the insulator or the superfluid, which are illustrated in Fig.~\ref{excite}.
For the insulator, the excitations are particle
or hole excitations above the background of the insulator with one particle per site. In contrast, for the superfluid, the excitations
are point-like vortices in the background of the Bose condensate. A semiclassical theory of gases of such excitations provides an essentially
complete description of the long-time correlations in the blue-shaded regions.

Let us now turn to the pink-shaded region of Fig.~\ref{qcfig}, separated from the blue-shaded regions by crossovers 
indicated by the dashed lines; the reader is referred to another review article \cite{physicstoday} for a detailed discussion
of the location and shape of these crossover lines. The defining characteristic of the pink-shaded `quantum critical' region is that
its dynamics is controlled by the CFT and its excitations. These excitations do not have a particle-like interpretation, they are
not amenable to an effective classical description, 
and they interact strongly (and universally) with each other. We are now faced with the challenge of describing the long-time
dynamics of this strongly interacting quantum critical regime.

The traditional renormalization group methods have been applied to the problem of quantum critical 
dynamics \cite{ssye,damle}. While a great deal of insight,
and some analytic results have been obtained, the results are not expected to be qualitatively accurate. We will shortly turn to
a description of the new holographic methods, and describe their promise in eventually solving this challenging problem.

It is helpful to focus on a specific observable characterizing the quantum critical dynamics in the pink-shaded region of Fig.~\ref{qcfig}. 
Let us choose the frequency-dependent
conductivity $\sigma (\omega)$: we endow the bosons with a (possibly fictitious) charge $q$, and examine the 
current response in the presence of an `electric' field coupling to the $q$ charge, which is oscillating at a frequency $\omega$. 
General arguments based on the properties of the CFT imply that \cite{damle}
\beq
\sigma (\omega)  = \frac{q^2}{\hbar} \Sigma \left( \frac{\hbar \omega}{k_B T} \right),
\eeq
where $\hbar$ is Planck's constant/($2\pi$), $k_B$ is Boltzmann's constant, and $\Sigma$ is an unknown, dimensionless universal
function of the dimensionless ratio of frequency to temperature. This structure follows from the fact that in $D=3$ the conductivity
is a dimensionless number when expressed in units of $Q^2/\hbar$; and the ability of the strong and universal 
interactions between the excitations of the CFT
to relax the electrical current. Each CFT in $D=3$ is characterized by its own function $\Sigma (\overline{\omega})$. 
We are now faced with the basic challenge: accurately determine the function $\Sigma (\overline{\omega})$ for the quantum critical 
point of the theory $\mathcal{S}_b$ in Eq.~(\ref{sb}) describing the superfluid-insulator transition. As we will see below, even simple questions
on the overall shape of this function remain unresolved.

The traditional field-theoretic methods (which are based on the renormalization group), as well as numerical studies, do work
accurately in determining the high frequency limit $\hbar \omega \gg k_B T$. Here, the correlations are essentially unchanged from
the ground state values, and so are amenable to a controlled analysis. The value of the number 
$\Sigma_\infty = \Sigma (\omega \rightarrow \infty )$ has been so determined \cite{mpaf1,mpaf2,fazio}, and there are no obstacles (in principle)
in refining this work to even more precise values.

How about determining the opposing limit \cite{longtime}, the value of the number $\Sigma_0 = \Sigma (\omega \rightarrow 0)$? 
This is a number characterizing the long-time response
at a non-zero temperature. So we have to extrapolate from our knowledge of the short-time dynamics to the long-time limit.
For systems with excitations which are weakly interacting particles, such extrapolations have traditionally been carried out by the
venerable Boltzmann equation, and its many descendants. Such methods are evidently not directly applicable to the
non-particle-like excitations of a CFT, which also have strong interactions.

Nevertheless, let us forge ahead and see what the Boltzmann approach teaches us about the qualitative shape of the
function $\Sigma$. Examining Fig.~\ref{excite}, we are immediately faced with a choice: do we approach the critical point
starting from the insulator or the superfluid? As we will now argue, a qualitatively different answer obtains from the two approaches.

\begin{figure}[htbp]
  \centering
  \includegraphics[width=4.5in]{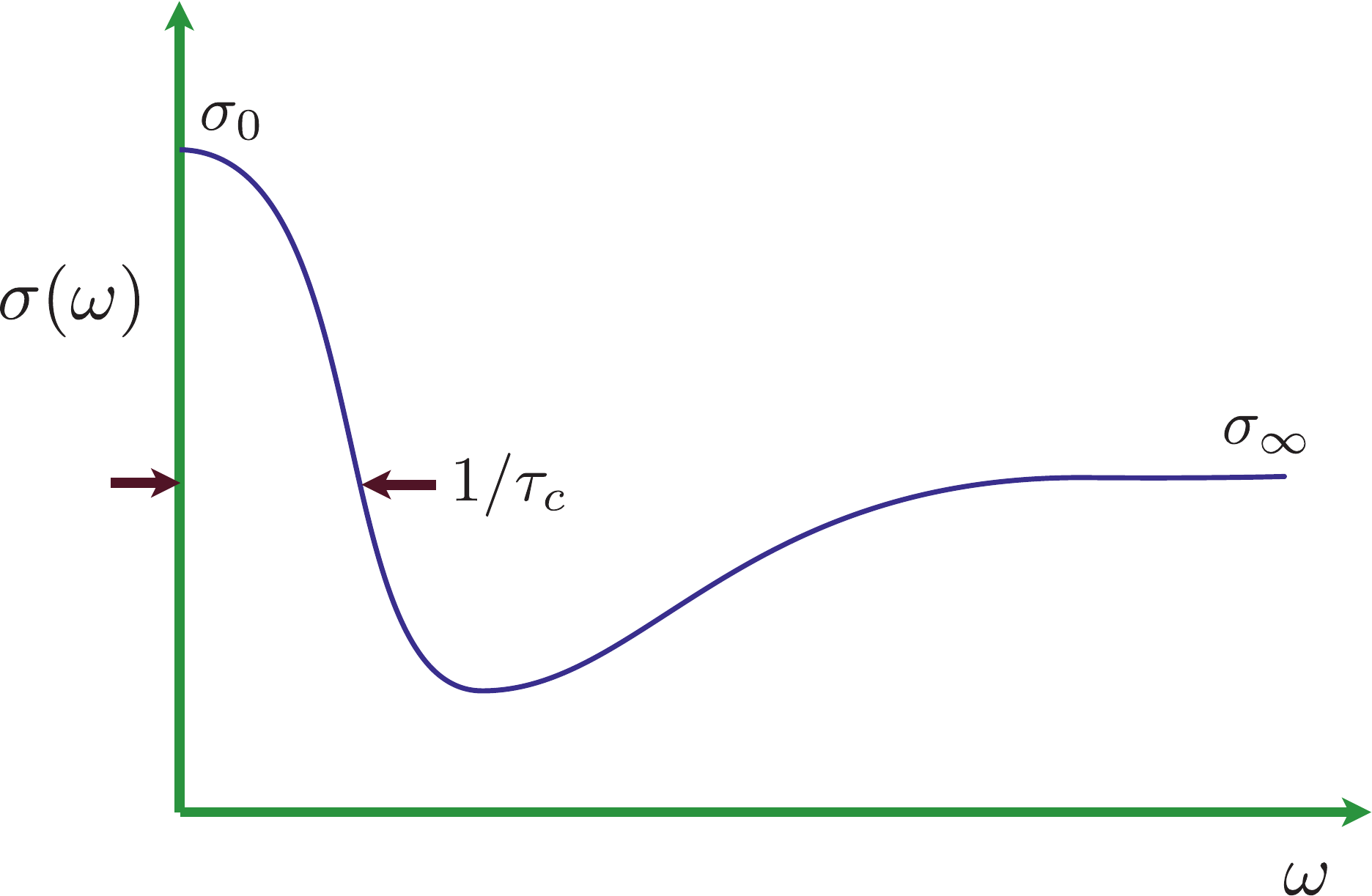}
  \caption{Expected form for the frequency-dependent conductivity of the CFT of the superfluid-insulator transition,
  using the Boltzmann picture applied to the particle excitations of the insulator \cite{damle}.}
  \label{particles}
\end{figure}

Let us begin by extrapolating the excitations of the insulator to the critical point. The insulator has particle and hole excitations,
and there is no difficulty in writing down a Boltzmann equation for these excitations \cite{damle}. 
In the simplest picture, thermally excited quasiparticles
undergo Brownian motion with mutual collisions, while drifting in the applied `electric' field $E$. The average velocity of these particles, $v$, therefore
obeys an equation like
\beq \frac{dv}{dt} + \frac{v}{\tau_c} = qE \label{brown}
\eeq
where $\tau_c$ is the mean time between collisions of the particles. If we extrapolate this picture to the critical point (without much justification),
we expect that this time would be determined by the only energy scale characterizing the CFT, which is $k_B T$,
and hence $\tau_c \sim \hbar/(k_B T)$.
Solving Eq.~(\ref{brown}, we then predict a ``Drude'' form for the frequency-dependent conductivity at
low frequencies, with
\beq \sigma (\omega) = \frac{\sigma_0}{1 - i \omega \tau_c}. \label{drude}
\eeq
Combining the real part of Eq.~(\ref{drude}) with our earlier considerations in the large $\omega$ limit, we can surmise 
a frequency dependent conductivity of the CFT with the qualitative form shown in Fig.~\ref{particles} \cite{damle}.
We reiterate that this form implicitly builds in the structure of the particle-like excitations of the insulator.

\begin{figure}[htbp]
  \centering
  \includegraphics[width=4.5in]{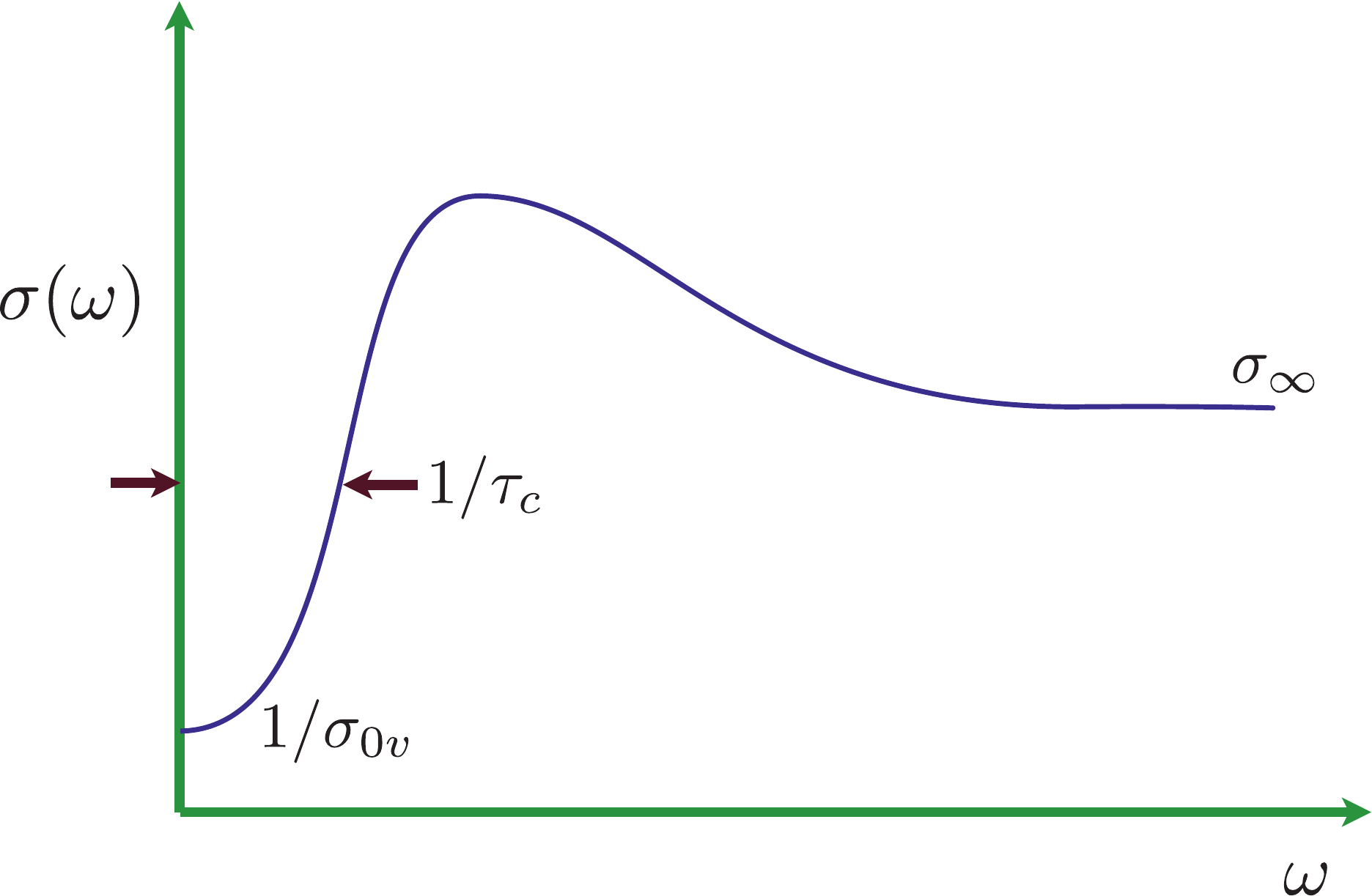}
  \caption{Expected form for the frequency-dependent conductivity of the CFT of the superfluid-insulator transition,
  using the Boltzmann picture applied to the vortex excitations of the superfluid. The ``d.c. conductivity'' of the vortex-like excitations
  is $\sigma_{0v}$.}
  \label{vortex}
\end{figure}

However, in principle, it should be equally valid to approach the critical point from the superfluid side, using its vortex-like excitations as the basic
degree of freedom. In two spatial dimensions, vortices can be specified by the location of their 
point-like center, and so can also be viewed as excitations
which are `particles'. We can, therefore, similarly write down a Boltzmann equation for the thermal and quantum dynamics
of these vortex excitations. We will refrain from doing so here, but note a basic property we will need: the physical conductivity
of the underlying boson model is equal to the resistivity of the vortex-like particles entering this Boltzmann treatment \cite{mpaf3}.
This can be deduced from the fact that a flow of vortices in a superfluid induces a voltage in the transverse direction.
From this we expect that the conductivity of the quantum critical point is roughly the inverse of the previous
prediction in Fig.~\ref{particles}; this result for the vortex picture of transport is shown in Fig.~\ref{vortex}.

We now have two qualitatively distinct predictions for the frequency-dependent function $\Sigma$, shown in Fig.~\ref{particles}
and~\ref{vortex}. It is a fair statement that for the CFT of the superfluid-insulator insulator transition described by the boson Hubbard
model, we do not know today which of these results, if either, is the correct one. The conventional $\epsilon$ and 
vector large $N$ expansions \cite{damle,ssqhe}
both have a built-in bias towards the excitations of the insulator, and so predict a form qualitatively similar to that in Fig.~\ref{particles}.
However, we are unable to say if these results should be believed for the physical case of interest. It is remarkable that this basic property
of one of the simplest quantum critical point in two spatial dimensions remains unresolved.

\subsection{Holographic analysis}
\label{sec:holo1}

Let us now change gears, and discuss what gauge-gravity duality can teach us about the questions at the 
end of the last section.

I will only be able to present a very short summary of the principles of gauge-gravity duality here; the reader is referred to the many
other reviews in the literature \cite{gubsertasi,hartnoll1,hartnoll2,herzog,horowitz,mcgreevy,tasi,statphys,greece}. 

The safest route is to begin with the solvable model, and then use physical arguments to generalize
to a wider class of CFTs, including perhaps those of physical interest to us. The canonical solvable model in $D=3$ spacetime dimensions
is a highly supersymmetric CFT now known as the ABJM model \cite{abjm}. In a particular `large $N$' limit, the low energy limit of 
this CFT is known to map onto a supersymmetric gravity theory in $D=4$ spacetime dimensions. We shall be interested here in the nature of
correlations of currents of globally conserved charges (analogous to the electrical current of the boson model above) in this theory.
Choosing a particular conserved current of the ABJM theory (it doesn't matter which one), its current correlations are
described by an especially simple and familiar theory of gravity and `electromagnetism' 
in $D=4$: the Einstein-Maxwell (EM) theory with a negative cosmological constant, with action \cite{mcgreevy}
\beq
\mathcal{S}_{EM} = \int d^4 x \left[ \frac{1}{2 \kappa^2} \left( R + \frac{6}{L^2} \right)
 - \frac{1}{4 e^2} F_{ab}F^{ab} \right].
 \label{Sem}
\eeq
Here $x \equiv (t, u, r)$ is a $4$-dimensional spacetime co-ordinate upon which the gravity theory is based; $r$ labels the $2$-dimensional
space upon which our CFT lives, $t$ is time, and $u$ is the new `emergent' co-ordinate of spacetime. 
The coupling $\kappa$ is related to Newton's gravitational constant by $\kappa^2 = 8 \pi G$.
The gravity theory is expressed in terms
of a metric $g$ and its Riemann curvature scalar $R$, while the U(1) gauge theory has `electromagnetic' flux $F_{ab}$;
we use small Latin letters for the spacetime co-ordinates of the $D+1$-dimensional spacetime.
Here, and henceforth, we have 
set the velocity of `light' to unity, $v=1$.

The physical significance of the gravity and the U(1) gauge field can be described after considering the vacuum solution
of the classical equations of motion associated with Eq.~\ref{Sem}. This solution has $F_{\mu\nu} = 0$, and the metric associated with
\beq
ds^2 = \left( \frac{L}{u} \right)^2 du^2 +  \left( \frac{u}{L} \right)^2 \left( - dt^2 + dr^2 \right)  \label{ads}
\eeq
This is the metric of the space of uniform negative curvature, known as 
AdS$_4$, and $L$ is the radius of curvature. It is evidently invariant under Lorentz transformations, and also under scale transformations
in (\ref{scal}) provided we choose
\beq
u \rightarrow u \, b; \label{scalu}
\eeq
furthermore, it is also invariant under
a suitable extension of the conformal transformation in (\ref{conform}).
This invariance is the crucial connection relating AdS$_4$ to conformal transformations, and to CFTs: the group of isometries of
AdS$_{D+1}$ is the same as the group of conformal transformations in $D$ spacetime dimensions. 
The transformation in (\ref{scalu}) also suggests a physical interpretation for the new co-ordinate $u$: it transforms like an energy/momentum scale,
and so can be viewed as the running energy/momentum scale for the renormalization group \cite{mcgreevy}. 
Thus the physics as $u \rightarrow \infty$ is the ultraviolet (UV) or short-distance/time physics, while the physics as $u \rightarrow 0$
is the infrared (IR) or long-distance/time physics. The gravity theory on AdS$_{D+1}$ maintains the complete `history' of the renormalization
group flow in the structure of the metric as a function of $u$.

We also note that the AdS space in Eq.~(\ref{ads}) has a boundary as $u \rightarrow \infty$. This boundary is just the $D$-dimensional
Lorentzian spacetime upon which our CFT of interest `lives'.

We are now in a position to describe the role of the U(1) gauge field $F_{ab}$ in Eq.~(\ref{Sem}). This is the field which
is `dual' to the conserved current of the CFT. Let us label the conserved current $J_\mu$; we use small Greek letters to 
represent the components of the $D$-dimensional Lorentzian spacetime. The conductivity is related to the two-point correlator of $J_\mu$.
To enable computation of this two-point correlator, let us include a source term coupling to the conserved current of the CFT:
\beq
\mathcal{S}_{\rm CFT} \rightarrow \mathcal{S}_{\rm CFT} - \int d^{D-1} r\, dt \, K_\mu J_\mu \label{cftK}
\eeq
Then a fundamental property of the gauge-gravity duality is that this source, $K_\mu$, is the limiting boundary value
of the vector potential $A_M$ associated with the U(1) flux $F_{ab}$ \cite{mcgreevy}:
\beq
A_\mu ( r, t, u \rightarrow \infty ) = K_\mu (r, t) \label{source}
\eeq
The complete prescription for computing the conductivity of the CFT using the holographic approach is as follows.
Solve the equations of motion of the gravitation theory subject to the constraint in Eq.~(\ref{source}), for an arbitrary spacetime
dependent $K_\mu (r,t)$. From these solutions, compute the functional dependence of the gravitational action on $K_\mu$.
This is precisely the functional dependence of the CFT action on $K_\mu$, and so correlators of $J_\mu$ are easily obtained by
taking functional derivatives of the action with respect to $K_\mu$.

Before we describe the results of such a computation, we need to discuss 2 additional points.

\begin{figure}[htbp]
  \centering
  \includegraphics[width=5in]{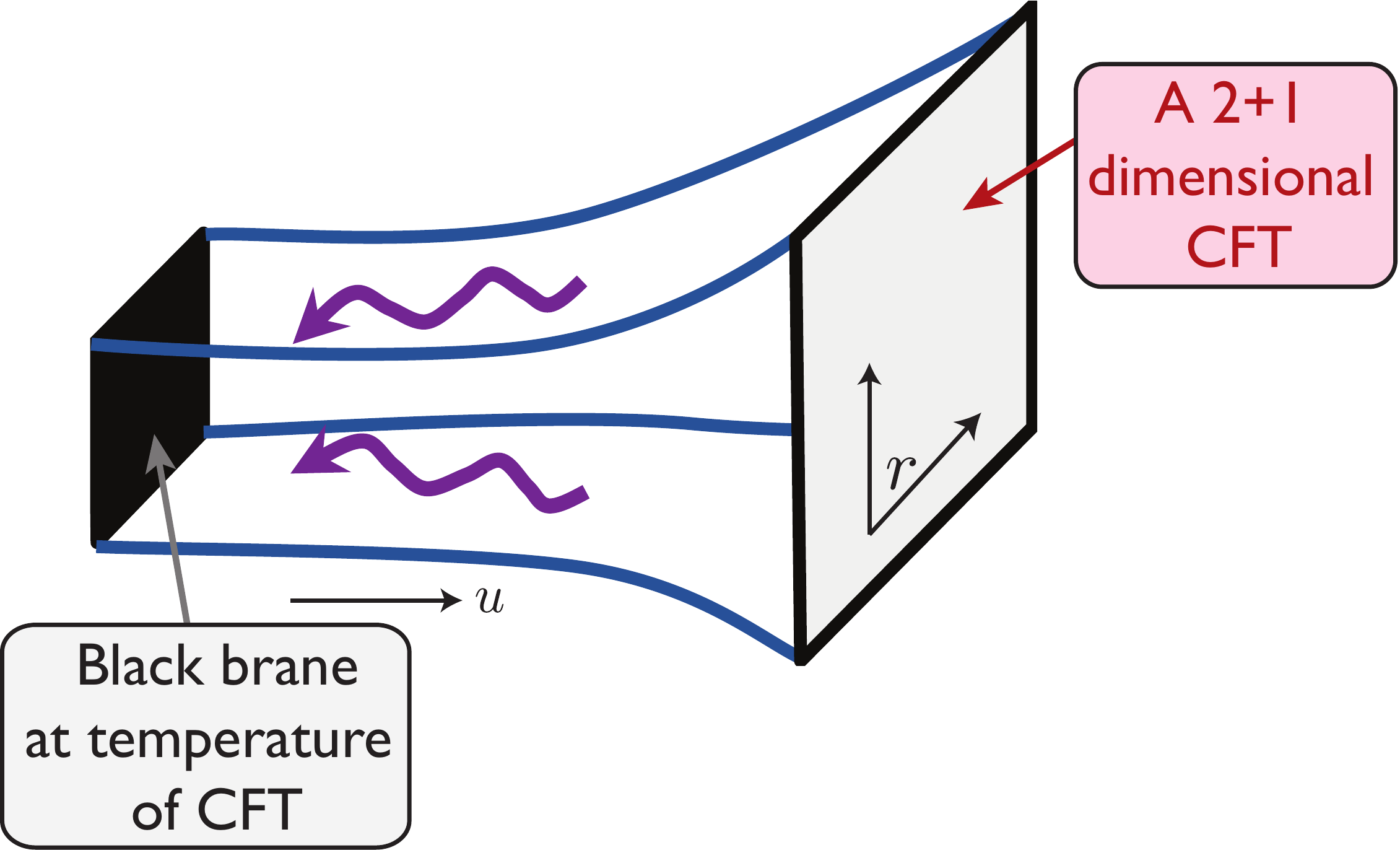}
  \caption{Schematic of the geometry of gauge-gravity duality for conformal quantum matter. 
  The gravity theory is defined on the bulk 4-dimensional spacetime
  (the time co-ordinate, $t$, is not shown). The CFT resides on the boundary at $u \rightarrow \infty$. Waves falling across the black brane
  at $u=R$ capture the dissipation of the CFT at non-zero temperature.}
  \label{black}
\end{figure}

First, the discussion above has implicitly assumed that we were at $T=0$. However, our most difficult questions about the CFT 
were at $T>0$, so it is essential we extend to non-zero temperatures. The key to this extension is to examine black hole solutions of the 
equations of motion of Eq.~(\ref{Sem}). These are analogous to the Schwarzschild solution, and take the form \cite{hartnoll2,mcgreevy}
\beq
ds^2 = \left( \frac{L}{u} \right)^2  \frac{du^2}{f(u)} +  \left( \frac{u}{L} \right)^2 \left( - f(u) dt^2 + dr^2 \right)  \label{bh}
\eeq
where
\beq
f(u) = 1 - \left( \frac{R}{u} \right)^3 \label{fu1}
\eeq
in $D=3$. Here $R$ is a parameter which labels the location of the black hole horizon. Also, strictly speaking, this is not a 
black hole, but a black brane: the horizon is spatially infinite, and extends across the flat two-dimensional $r$ space.
As shown by Hawking \cite{hawking}, the black brane horizon must have a temperature, $T$, and in the present duality we can identify
this temperature with the temperature, $T$, of the CFT. This Hawking temperature is most simply computed by analytically
continuing the metric in Eq.~(\ref{bh}) to imaginary time, and demanding that the resulting space be periodic in the imaginary
time direction with period $\hbar / (k_B T)$; such a computation yields
\beq
k_B T = \frac{3 \hbar v R}{4 \pi L^2} \label{ht}
\eeq
where we have momentarily reinserted the factor of $v$; this can be viewed as an equation which fixes the value of $R$.
This black-brane is a powerful feature of the present holographic approach. It enables us to see how dissipative 
and relaxational processes of the CFT at non-zero temperature have a natural interpretation in the gravitational theory.
As illustrated in Fig.~\ref{black}, waves propagating in the $D+1$ dimensional space get damped because they
lose energy across the black-brane horizon: it is this damping which Eq.~(\ref{source}) eventually relates to the dissipative
transport co-efficients of the CFT \cite{kss}.

Second, as we noted earlier, the action in Eq.~(\ref{Sem}) was obtained for the special CFT described by ABJM.
It also describes a very wide class of related supersymmetric CFTs, but this does not include the CFT associated with 
the boson Hubbard model in Eq.~(\ref{sb}). To extend to this wider class, we will use the spirit of effective field
theory, applied to the holographic method as proposed recently in Ref.~\onlinecite{ajay}. 
We can also view Eq.~(\ref{Sem}) as the simplest effective action
for the metric and the U(1) gauge field, invariant under the underlying symmetries, in which all fields
are expanded to include 2 gradients of spacetime co-ordinates. Let us assume this is a reasonable starting point, and extend
this action to include 4 spacetime gradients. As we are only interested in linear response to the source $K_\mu$ in Eq.~(\ref{cftK}),
we can restrict this extension to include only 2 powers of $A_M$. With these conditions, it turns out only one
additional term is allowed, modulo allowed co-ordinate transformations, and the extended action is \cite{ajay}
\beq
\mathcal{S} = \mathcal{S}_{EM} + \int d^4 x \sqrt{-g} \, \frac{\gamma L^2}{e^2} \, C_{abcd} F^{ab} F^{cd}, \label{weyl}
\eeq
where $C_{abcd}$ is the Weyl curvature tensor. 
Eq.~(\ref{weyl}) is our final theory for a very wide class of CFTs, which we hope applies also to the CFT of the boson Hubbard
model with reasonable accuracy. This theory depends upon two-dimensionless parameters: $e$ and $\gamma$. 
However, both parameters can, in principle, be fixed by matching to the short-time, or $\omega \rightarrow \infty$, limit of the correlators
of the CFT of interest; $e$ is related to $\Sigma_\infty$, while $\gamma$ connects to a 3-point correlator
of the current $J_\mu$ and the stress-energy tensor \cite{ajay}. 

Let us summarize our proposal for applying gauge-gravity duality to the CFT realized in the boson Hubbard model.
We use effective field theory to motivate the effective action of the $D+1$ dimensional gravity theory in Eq.~(\ref{weyl}).
Conventional field-theoretic methods can be used to compute characteristics of the CFT at $T=0$, and
these can be matched to the gravitational theory to fix the values of parameters.
Then we extrapolate to the long time limit at $T>0$ using the
solution of the gravity equations of motion in the black brane background, as describe below Eq.~(\ref{source}).
It is interesting that this procedure parallels that used for Boltzmann-like equations: use properties of the quasiparticles at
$T=0$ to compute their scattering cross-section, so determine the collision term, and then solve the Boltzmann equation
to extrapolate to the long time limit at $T>0$. Evidently, the $D+1$ dimensional gravity theory has taken the place
of the Boltzmann equation for CFTs without quasiparticle excitations.

\begin{figure}[htbp]
  \centering
  \includegraphics[width=6in]{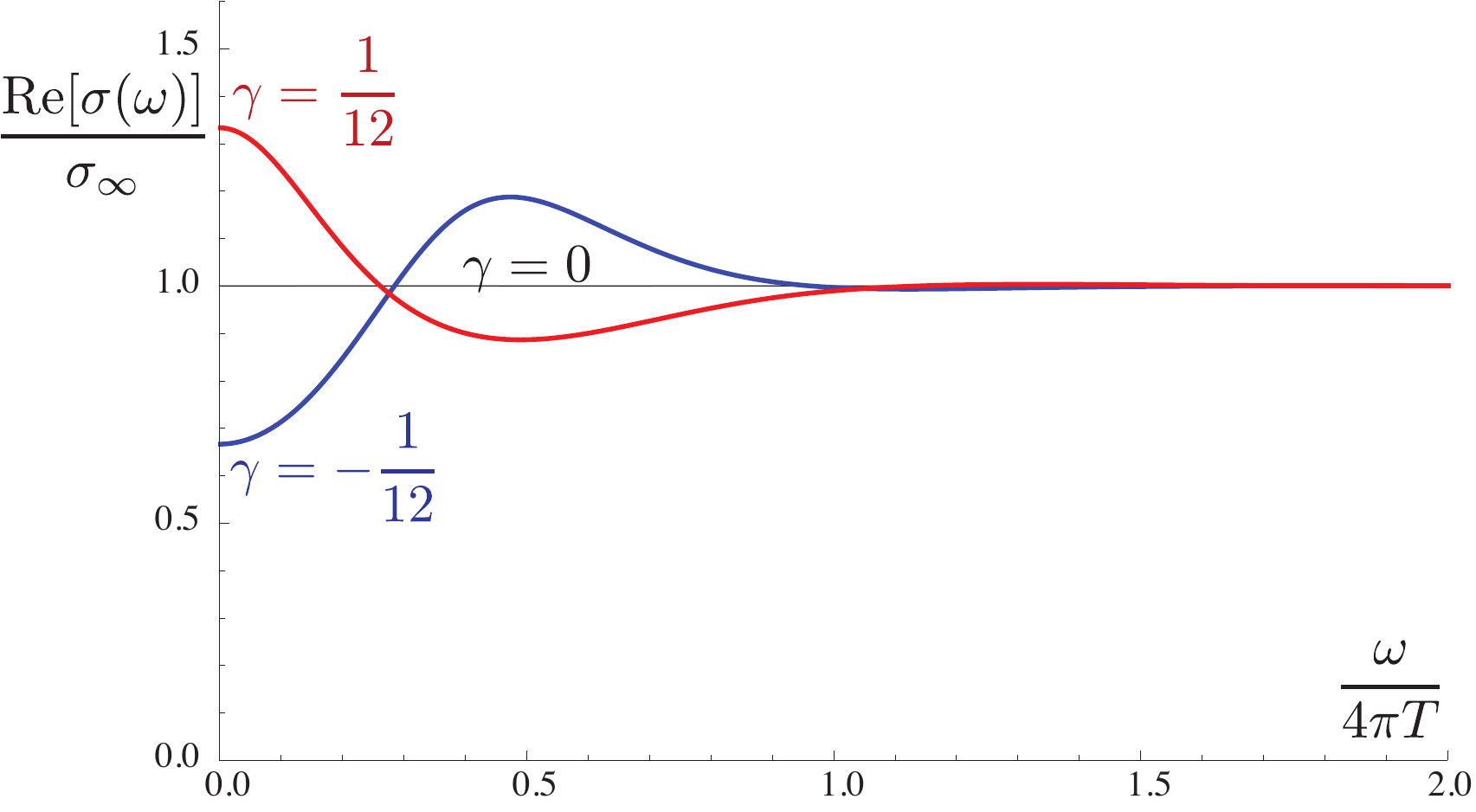}
  \caption{Predictions \cite{ajay} for the frequency-dependent conductivity $\sigma (\omega)$ of a CFT in 2+1 dimensions from the holographic
  gravity theory in Eq.~(\ref{weyl}). The results
  shown depend only upon the parameter $\gamma$ which is restricted by stability requirements to the range $|\gamma | < 1/12$.
  The $\gamma=0$ case has a particle-vortex self-duality and applies to special supersymmetric models like the ABJM theory.}
  \label{ajay}
\end{figure}

We are now ready to describe the results of this method in the computation of $\sigma (\omega)$. The results depend upon
the values of $e$ and $\gamma$. However, the dependence on $e$ can be scaled out by examining the ratio
$\sigma (\omega) /\sigma_\infty$. Furthermore, stability of the holographic theory
requires that $|\gamma| < 1/12$ \cite{ajay}. Explicit results for this range of $\gamma$ are shown in Fig.~\ref{ajay}. 
We see from Fig.~\ref{ajay} that the change in $\sigma (\omega) $
from the $\gamma = 0$ result to the maximal allowed values of $\gamma$ is quite 
limited: this stability is very encouraging, and is a partial a posteriori justification for the
validity of the gradient expansion applied to the holographic theory. 

A remarkable feature of Fig.~\ref{ajay} is that the results correspond very neatly to the conductivities surmised from the Boltzmann
equation in Figs.~\ref{particles} and~\ref{vortex}: the $\gamma>0$ case is similar to the particle Boltzmann sketch in Fig.~\ref{particles},
while the $\gamma < 0$ case is similar to the vortex case in Fig.~\ref{vortex}. Thus it is the sign of $\gamma$ which determines
whether a given CFT is more accurately described by the particle-like or vortex-like excitations.
We do not yet know the value of $\gamma$ for the CFT of the boson Hubbard model, but the route is now open to its determination
by the strategy we have outlined above.

We close by noting a curious feature of Fig.~\ref{ajay}. The $\gamma = 0$ case yields a frequency-independent $\sigma (\omega)$.
Recall that $\gamma =0$ corresponded to the supersymmetric ABJM-like theories. These special theories have a {\em self\/}-duality
under the particle-vortex transformation which accounts for this novel feature \cite{m2cft}.

\section{Compressible quantum matter}
\label{sec:nfl}

\subsection{Condensed matter analysis}
\label{sec:cond2}

We begin by defining compressible states, in a form applicable to both condensed matter
and holographic theories \cite{liza}.
\begin{itemize}
\item Consider a continuum, translationally-invariant 
quantum system with a globally conserved charge $\mathcal{Q}$
{\em i.e.\/} $\mathcal{Q}$ commutes with the Hamiltonian $H$. Couple the Hamiltonian
to a chemical potential, $\mu$, which is conjugate to $\mathcal{Q}$: so the Hamiltonian
changes to $H - \mu \mathcal{Q}$. The ground state of this modified Hamiltonian
is compressible if $\langle \mathcal{Q} \rangle$ changes smoothly as a function of $\mu$, 
with $d\langle \mathcal{Q} \rangle/d\mu$ non-zero.
\end{itemize}
Note that our definition of compressibility refers to the ground state, and so we need $T=0$.
Also, as noted in Section~\ref{sec:intro}, we will restrict our attention to compressible states
in dimensions greater than unity ($D>2$).

Remarkably, it turns out there are only a few known states which satisfy these seemingly innocuous
requirements. The states are:
\begin{enumerate}
\item \underline{Solids:} Translational symmetry is broken, and the matter `crystallizes' into a periodic 
arrangement. Changing $\mu$ changes the period of the lattice, allowing a continuous variation in 
$\langle \mathcal{Q} \rangle$.
\item \underline{Superfluids:} Here the global U(1) symmetry associated with conservation of $\mathcal{Q}$
is spontaneously broken. These are gapless states which readily accept new particles into the Bose condensate,
allowing a smooth variation in the density $\langle \mathcal{Q} \rangle$.
\item \underline{Fermi liquids:} The simplest example of a Fermi liquid is the non-interacting Fermi gas. The fermionic particles
occupy momentum eigenstates, with momenta inside a {\em Fermi surface\/} which separates the occupied and empty states.
This picture generalizes to interacting particles, to all orders in perturbation theory, as shown elegantly by Landau.
In a Landau Fermi liquid, there is a sharp Fermi surface, and the only low energy excitations are sharp quasiparticles
(renormalized from the bare fermions) in the vicinity of the Fermi surface. These quasiparticles 
carry opposite $\mathcal{Q}$ charges on either side of the Fermi surface. Changing $\mu$ changes the location of the Fermi 
surfaces, and the newly occupied states allow a smooth variation in $\langle \mathcal{Q} \rangle$.
\end{enumerate}
Apart from mild variations which combine features of such states, these are the only examples of compressible
quantum states in traditional condensed matter physics.

The past decades have seen intensive search for other `exotic'  compressible states of condensed matter.
As reviewed more completely in Ref.~\onlinecite{physicstoday}, this search has been motivated by the 
ubiquitous appearance of `strange metal' behavior in a variety of correlated electron compounds, including
the high temperature superconductors. From such studies a new class of compressible states appears to have
emerged, which we refer to here generically as `non-Fermi liquids'. 
Many recent studies \cite{book,motfish,leen,metnem,mross,metsdw,sdwtransport} have emphasized the strong-coupling
nature of the theory of non-Fermi liquids in two spatial dimensions ($D=3$), 
and noted that important questions remain unresolved.

Let us describe one of the simplest examples of a compressible non-Fermi liquid state. This was developed
as a theory of a possible spin liquid (SL) state in frustrated quantum antiferromagnets \cite{baskaran,mot1,mot2,senthilmott,leem},
and a complete derivation appears in the Supplementary Material.
The primary degrees of freedom
are fermionic `spinons' $f_\alpha$, with $\alpha = \uparrow, \downarrow$ a spin label.
These are strongly coupled to an `emergent' U(1) gauge field $B_\mu$. Note that this gauge field is not
to be confused with Maxwell gauge field associated with electromagnetism, whose fluctuations are ignored in all our
analyses. It should also not be confused with the gauge field $A_a$ in Section~\ref{sec:holo1}, which resides
in the extended $(D+1)$-dimensional space. Rather, $B_\mu$ is a degree of freedom which emerges from the 
dynamics of the antiferromagnet, and encodes the complex quantum entanglement between valence bonds in the spin liquid.
The action for this non-Fermi liquid state is simple (see Supplementary Material)
\beq
\mathcal{S}_{{\rm SL}} = \int d^2 r dt \left[ f^\dagger_\alpha \left(  i\frac{\partial}{\partial t} +  B_t - \varepsilon_F 
- \frac{1}{2m} (\nabla_r - i B_r )^2 \right) f_\alpha - B_t \mathcal{N} \, \right] \label{ssl}
\eeq
where both $f_\alpha$ and $B_\mu$ are fluctuating functions of $r$ and $t$, and the remaining parameters are constants. 
The Fermi energy $\varepsilon_F$ controls the
value of the fermion density $ \langle f_\alpha f_\alpha \rangle$, but we are not free to choose its value. This density is coupled to a 
fluctuating U(1) gauge field which mediates long-range `Coulomb' interactions, 
and so just as in the jellium model of the electron gas, 
stability of the system requires net neutrality in the gauge charge. We have thus included a background
neutralizing charge density $\mathcal{N}$, and the value of $\varepsilon_F$ must be chosen so that
\beq
\langle f_\alpha^\dagger f_\alpha \rangle = \mathcal{N}. \label{back}
\eeq
Thus the system is not compressible with respect to the charge $f_\alpha^\dagger f_\alpha$; moreover this charge is not associated with
a global conservation law, but with a gauge invariance. Instead, the quantity realizing the globally conserved charge $\mathcal{Q}$ associated
with compressibility is 
the {\em spin density}. The spin density has 3 components, and let us arbitrarily choose the $z$ component, and so
\beq
\mathcal{Q} = f_\alpha^\dagger \sigma^z_{\alpha\beta} f_\beta,
\eeq
where $\sigma^z$ is a Pauli matrix. The ``chemical potential'' coupling to $\mathcal{Q}$ is the Zeeman coupling of an applied
magnetic field. The compressibility of the SL state implies that its magnetization can be varied smoothly as a function of the applied
magnetic field, and the magnetic susceptibility is non-zero at $T=0$ \cite{mot1}.

Like Landau Fermi liquids, a non-Fermi liquid SL state also has Fermi surfaces, but the character of the fermionic excitations
near the Fermi surface is quite different. Formally, for interacting electron systems, 
the Fermi surface is defined by the momentum $k=k_F$ 
for which there is a zero in the inverse fermion Greens function:
\beq
G_f^{-1} (k=k_F, \omega = 0) = 0 \label{fs}
\eeq

Landau's Fermi liquid theory also predicts a specific singularity in the fermion Green's function near the Fermi surface:
this singularity reflects the fact that the inverse-lifetime of the quasiparticles vanishes as the square of their distance from the Fermi surface.
The energy of the quasiparticles is parametrically larger: it is linear in the distance from the Fermi surface, and so the quasiparticles
are well-defined excitations.

\begin{figure}[htbp]
  \centering
  \includegraphics[width=3.5in]{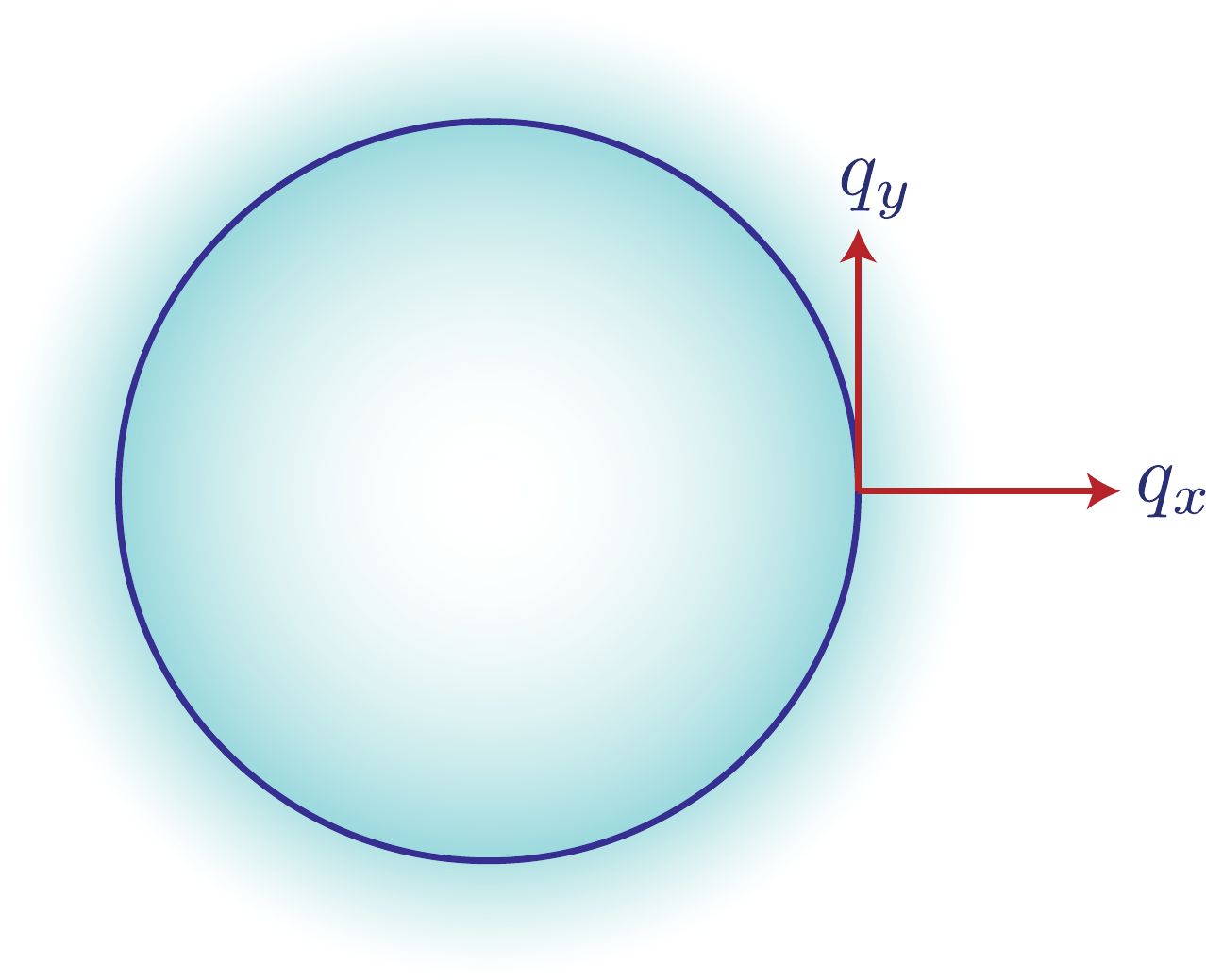}
  \caption{Fermi surface of a non-Fermi liquid. The ``fuzzy'' shading is used to indicate the broad spectral functions of the 
  fermionic excitations. Note that despite this fuzziness, the position of the Fermi surface in momentum space is sharply defined.}
  \label{fsf}
\end{figure}

In the known non-Fermi liquids, the quasiparticles do not remain well-defined excitations away from the Fermi surface: they
are strongly scattered by the low energy modes the U(1) gauge field. This is reflected in the singularity of the fermion
Green's function near the Fermi surface, which has been argued to have the generic form \cite{metnem}
\beq
G_f^{-1} (k, \omega) = q^{1-\eta} \, F (\omega/q^{z/2}) \label{fsnfl}
\eeq
where we are expanding in the vicinity of the Fermi surface point at $(k_F, 0)$, with 
$q_x = k_x - k_F$, $q_y = k_y$, and $q = q_x + q_y^2$ (see Fig.~\ref{fsf}). 
The universal scaling function $F$ is a complex-valued scaling function
which determines the spectral density of the fermionic excitations. The parameters $\eta$ and $z$ are
critical exponents whose values can be estimated in various expansion methods \cite{metnem,mross}. However, the theory of these
fermionic and gauge modes is strongly coupled \cite{leen,metnem} (in $D=3$), and so the reliability of these estimates is unknown.

We now make a few more remarks about the Fermi surfaces encountered so far:\\
({\em i\/}) The Fermi surface of non-Fermi liquids
is sharply defined, even though the fermionic quasiparticles are not: the value of $k_F$ is precisely defined by Eq.~(\ref{fs}),
and the singularity of the scaling function in Eq.~(\ref{fsnfl}) on the Fermi surface.\\
({\em ii\/}) The above results are expected to apply
also to fermions coupled to non-Abelian gauge fields, or to order parameter fluctuations near a symmetry-breaking
quantum phase transition. Unlike the familiar case at zero density, there does not
appear to be any fundamental difference between Abelian and non-Abelian gauge fields. \\
({\em iii\/}) There is a crucial relation, known as
the Luttinger relation, which constrains the total area enclosed by Fermi surfaces to the values of conserved charges.
Specifically, let us consider a compressible state with Fermi surfaces labeled by $\ell$: each Fermi surface is associated with the singularity
(\ref{fs}) in the Green's functions of fermions carrying global charge $q_\ell$ under symmetry associated with $\mathcal{Q}$, 
and encloses area $\mathcal{A}_\ell$. Then the Luttinger relation is \cite{powell,piers}
\beq
\sum_\ell q_\ell \mathcal{A}_\ell = 4 \pi^2 \langle \mathcal{Q} \rangle. \label{lut}
\eeq
The fermionic excitations near the Fermi surface can also carry charges associated with fluctuating gauge fields, as in Eq.~(\ref{ssl}).
If none of the Fermi surfaces have gauge charges, we usually obtain a Fermi liquid. If we have Fermi surfaces with and without
gauge charges co-existing, we obtain a compressible state labeled as a {\bf fractionalized Fermi liquid} \cite{ffl1,ffl2}.

A Luttinger relation like Eq.~(\ref{lut}) applies for each conserved U(1) charge $\mathcal{Q}$ associated with a compressible 
state. We also note \cite{liza} that for the case of an Abelian gauge field (as in Eq.~(\ref{ssl})), there is an additional requirement for
global neutrality in the gauge charge: in this case, there is additional Luttinger relation for the Fermi surfaces
carrying the gauge charges, equating the sum of their areas to the background charge (which is $\mathcal{N}$ in Eqs.~(\ref{ssl}) and
(\ref{back})).

Our discussion of compressible states of condensed matter now suggests a conjecture \cite{liza}:
\begin{itemize}
\item
All compressible states which preserve translational and global U(1) symmetries must have
Fermi surfaces, but they are not necessarily
Fermi liquids. The areas enclosed by these Fermi surfaces obey a Luttinger relation for each conserved charge. 
\end{itemize} 
Note that we have not imposed any requirement that our underlying Hamiltonian have any fermionic
degrees of freedom. It could be defined solely in terms of bosons, or have both fermions and bosons; 
recent examples of boson-only models are in Refs.~\onlinecite{motfish,motfish2,motfish3}.
There are fermionic excitations near each Fermi surface, and these fermions could be fractions or composites
of the underlying particles.

The rationale behind this conjecture is that (in $D-1$ spatial dimensions) we need a surface of dimension $D-2$ of zero energy excitations
to have adequate phase space to allow $\langle \mathcal{Q} \rangle$ to vary smoothly as a function of the chemical
potential. In $D>2$, bosons can't generically have such a surface of zero energy excitations: there will be negative energy 
states on one side of this surface, rendering the system unstable to Bose condensation. In contrast, fermions are allowed
to have such surfaces with a singularity as in Eq.~(\ref{fs}), as is evident already from the free fermion theory: the negative energies
now represent hole-like excitations of the fermions.

We are finally in a position to state the challenge to the holographic approach to condensed matter physics. Can gauge-gravity duality provide
a classification of possible states of compressible quantum matter? Clearly, any such classifications should include the
familiar phases: solids, superfluids, and Fermi liquids. Are there additional states, and do they correspond
to the non-Fermi liquid states with Fermi surfaces described above? If so, we can hope that they will provide a
new perspective on their strong-coupling physics. Or is the conjecture incorrect, and are there entirely new types of compressible
quantum states which have been overlooked in condensed matter studies?

\subsection{Holographic analysis}
\label{sec:holo2}

The study of compressible states using holography is the focus of much current 
research \cite{gubsertasi,hartnoll1,hartnoll2,herzog,horowitz,mcgreevy,tasi,statphys,greece}. This research 
does not yet have definitive answers
to the questions posed at the end of the previous section. However, rapid progress has been made recently, and here I will give
my perspective on the current state of the theory. This research promises to eventually achieve a holographic classification of the 
possible states of compressible quantum matter. Such a classification 
will be useful for condensed matter applications, especially in two spatial dimensions, 
regardless of the artificial nature of the microscopic degrees
of freedom used to realize the compressible phases. 

The basic starting point is to extend the Einstein-Maxwell
theory $\mathcal{S}_{EM}$ in Eq.~(\ref{Sem}) which we used in the conformal case to a situation with non-zero density. Given
the relationship in Eq.~(\ref{source}), we can turn on a finite density by including a chemical potential, in which case we now have
the boundary condition
\beq
A_t ( r, t, u \rightarrow \infty ) = \mu , \label{sourcemu}
\eeq
while the other components of the gauge field vanish in the corresponding limit near the boundary. We will now work to all orders in $\mu$, rather than the linear response in the source term in Section~\ref{sec:holo1}. 

\begin{figure}[htbp]
  \centering
  \includegraphics[width=4in]{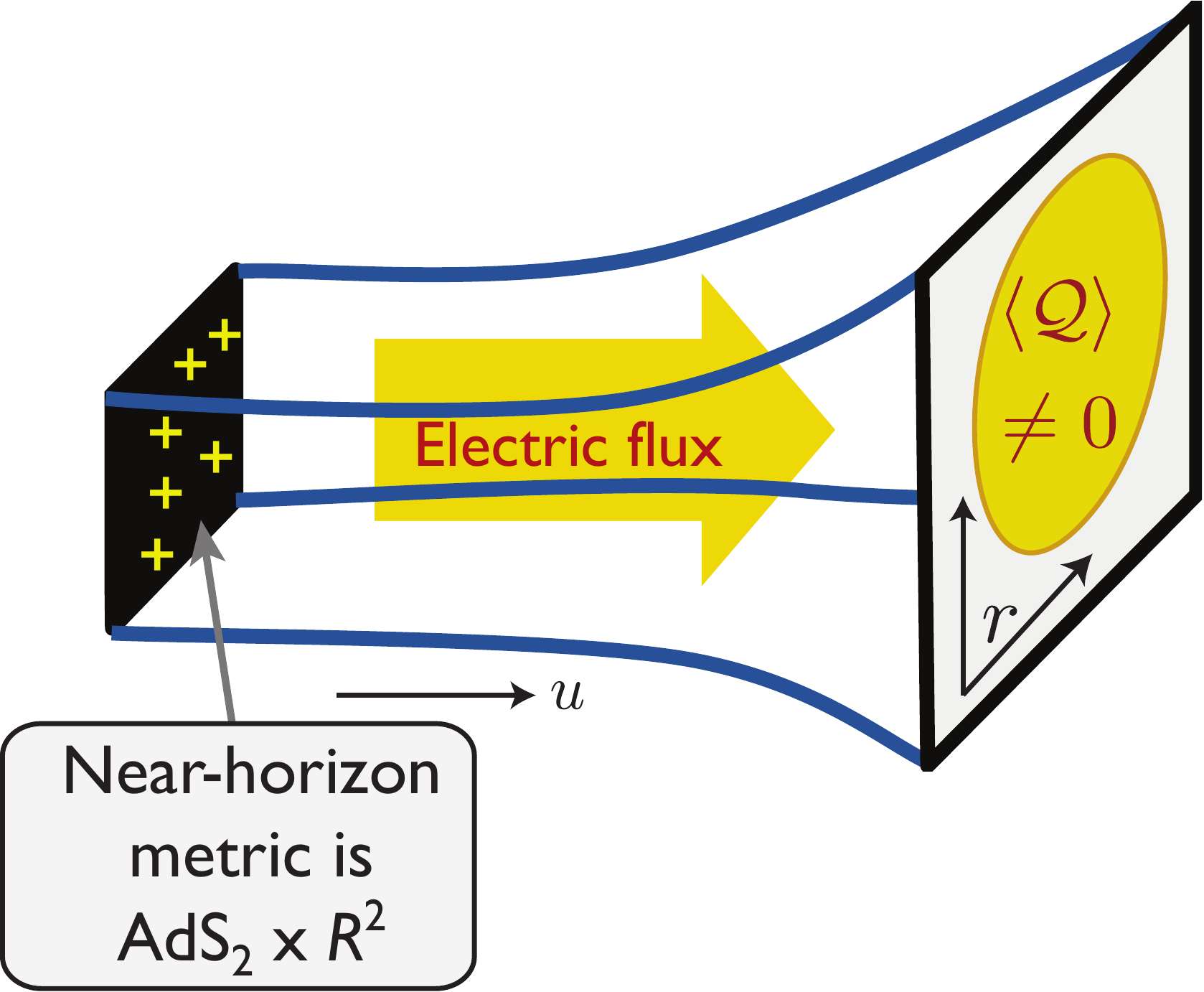}
  \caption{Schematic (adapted from Ref.~\onlinecite{hartnoll1}) 
  of the proposed geometry of gauge-gravity duality for compressible quantum matter, generalizing
  Fig.~\ref{black} to a AdS$_4$-Reissner-Nordstr\"om black brane. 
  The non-zero density $\langle \mathcal{Q} \rangle$ on the boundary sources an electric flux
  which extends into the bulk to the horizon. Consequently, there must be compensating charges sourcing
  the electric flux `behind' the horizon.
  }
  \label{rnbh}
\end{figure}

In the simplest (and frequently used) approach, we make no further change. We solve the equations of motion 
of the action in Eq.~(\ref{Sem}) subject to the boundary condition in Eq.~(\ref{sourcemu}) for the gauge field. Such a 
solution is of the Reissner-Nordstr\"om form \cite{rob}: the metric remains as in Eq.~(\ref{bh}), but the function $f(u)$
in Eq.~(\ref{fu1}) is replaced by
\beq 
f(u) = 1  + \frac{L^2 \kappa^2 \mu^2}{2 e^2 R^2}\left(\frac{R}{u} \right)^4 - \left(1 + \frac{L^2 \kappa^2 \mu^2}{2 e^2 R^2} \right)\left(\frac{R}{u} \right)^3 
\label{fu2}
\eeq
Again $R$ is the position of the horizon, as it can be checked that $f(R)=0$. In addition, we have a non-zero gauge field given by
\beq
A_t (u) = \mu \left( 1 - \frac{R}{u} \right), \label{atu}
\eeq
which obeys Eq.~(\ref{sourcemu}), and vanishes on the horizon. As above Eq.~(\ref{ht}), we can compute the temperature of the horizon,
and now find that it equals
\beq
k_B T = \frac{3 \hbar v R}{4 \pi L^2} \left( 1 - \frac{L^2 \mu^2 \kappa^2}{6 e^2 R^2} \right); \label{Tmu}
\eeq
as in Eq.~(\ref{ht}), this equation fixes $R$ in terms of physical parameters.
A sketch of this geometry, generalizing Fig.~\ref{black}, is shown in Fig.~\ref{rnbh}. 
Note that the total `electric' flux
in the bulk is conserved between the horizon and the boundary at $u=\infty$: this follows from the potential in Eq.~(\ref{atu}), which shows
that the electric field $\sim 1/u^2$, and the metric in Eq.~(\ref{bh}), which shows that the surface area $\sim u^2$.

This gravitational solution can now be applied to examine physical properties of the quantum system on the
boundary as a function of the temperature, $T$, and the chemical potential, $\mu$. For $T \gg \mu$ this was examined in 
Ref.~\onlinecite{markus}, and successfully used to develop a general theory of thermoelectric transport in compressible
quantum systems near quantum critical points. 
We will not review this here, as our primary interest is in the
limit $T \ll \mu$ for non-superfluid states. I also note in passing the interesting recent work of Bhattacharya {\em et al.} \cite{shiraz} 
which has obtained new insights on the low temperature hydrodynamics of superfluids using the dual gravity approach.

A notable feature of the $T \rightarrow 0$ limit of the above Reissner-Nordstr\"om solution is apparent 
from Eq.~(\ref{Tmu}): the horizon radius, $R$ remains finite when $\mu$ is non-zero, with
\beq
R = \frac{L \kappa}{\sqrt{6} e} \mu \quad;\quad T=0. \label{R0}
\eeq
This is different from the conformal case of Section~\ref{sec:holo1}, where the horizon was absent at $T=0$.
An immediate consequence is that our quantum system on the boundary, has a non-zero entropy density
in its ground state, as can be verified by a computation of the free energy of the gravitational solution.
The free energy computation \cite{hartnoll2} also shows that the compressibility of the ground state is finite, where the charge density
is a smooth function of $\mu$:
\beq
\langle \mathcal{Q} \rangle = \frac{\kappa}{\sqrt{6} e^3 L} \mu^2 \quad;\quad T=0. \label{comp}
\eeq
So we have succeeded in obtaining a phase of compressible quantum matter, but it has the unacceptable feature
of a ground state degeneracy which increases exponentially with system size.

However, we will not discard this solution right away: let us examine some of its properties more carefully
and see if a physical interpretation emerges. This will also help point the way towards improving this holographic
description. 

We examine the form of the metric more closely; using the relationship (\ref{R0}) in
Eq.~(\ref{fu2}), we find
\beq
f(u) = \left(1 - \frac{R}{u} \right)^2 \left( 1 + \frac{2 R}{u} + \frac{3 R^2}{u^2} \right) \quad ; \quad T=0 \label{fu3}
\eeq
Note that $f(u)$ has a double zero at the horizon $u=R$, unlike the conventional single zero in the conformal case in Eq.~(\ref{fu1}).
Such a horizon is known as {\em extremal}, and this feature is linked to its anomalous properties. Defining a co-ordinate which 
is zero at the horizon, $\tilde{u} = u - R$, and expanding to the lowest non-vanishing order in $\tilde{u}$, the metric
in Eq.~(\ref{bh}) becomes
\beq
ds^2 = \frac{1}{6} \left( \frac{L}{R} \right)^2  \frac{d \tilde{u}^2}{\tilde{u}^2} -   6 \left( \frac{R}{L} \right)^2 \tilde{u}^2 dt^2 
+ \left( \frac{R}{L} \right)^2 dr^2 . \label{ads2}
\eeq
The notable feature of this metric is its separation into the first two terms dependent only upon $t$ and $\tilde{u}$,
and the last term dependent only upon the spatial co-ordinate. This means that spacetime has factorized into
a spatial $R^2$, and a curved space in $\tilde{u}$ and $t$. By comparison with Eq.~(\ref{ads}), the reader will recognize
that the latter space is AdS$_2$, and so we conclude that the near-horizon spacetime of a Reissner-Nordstr\"om black brane
is AdS$_2 \times R^2$. This factorized form of the metric has strong consequences for all the low energy properties 
of the compressible quantum system on the boundary.

Let us begin by interpreting the AdS$_2$ factor. What kind of quantum system does this describe?
As I have reviewed in more detail elsewhere \cite{tasi,statphys,ssffl}, it describes the physics of 
a single quantum impurity universally coupled to a CFT, analogous to the overscreened
fixed points of multichannel Kondo problems \cite{pgks}, or a vacancy in a two-dimensional insulating 
antiferromagnet at a magnetic  ordering critical point \cite{science}. There is a conformal structure to the correlations near the 
impurity, and the ground state has a finite entropy in the limit $T \rightarrow 0$ \cite{pgks,pgs}. All these features are physical and robust,
and potentially realizable in generic experimental systems. Supersymmetric generalizations of such quantum impurity systems
can be solved \cite{shamit} by gauge-gravity duality, and explicitly reveal an AdS$_2$ metric in the low
energy physics.

We are now ready to interpret AdS$_2 \times R^2$. The factorized form of the metric means that
we can view the two-dimensional quantum system as consisting of an infinite number of 
quantum impurity problems, one at each position in space. Because there is an ultraviolet cutoff above which the 
AdS$_2$ factorization does not hold, there is a similar cutoff in spatial separation below which
the quantum impurities remain coupled. Thus the low energy physics is controlled by a {\em finite density\/}
of {\em independent\/} quantum impurities. Each quantum impurity is coupled to a gapless environment, which
is presumably a mean-field description of the other quantum impurities. Many condensed matter theorists will immediately
recognize the similarity of this picture to that of dynamical mean field theory (DMFT) \cite{dmft}.  Such large dimension/co-ordination number
approximations have also been applied to the vicinity of magnetic ordering transitions, and descriptions of compressible
non-Fermi liquid states have been obtained \cite{sy,pg,pgs,bgg,si1,si2}. Such states now have a non-zero ground state entropy 
{\em density\/} \cite{pgks,pgs},
a consequence of adding up the entropy of the finite density of decoupled quantum impurity problems. It was argued \cite{ssffl}
that these compressible critical states, obtained in the limit of large spatial dimension or lattice co-ordination number, which 
provide a specific microscopic realization of the physics of AdS$_2 \times R^2$. The holographic theory shares the feature of having
a non-zero ground state entropy density, and we will now see that the structure of correlations of local operators also match.

Correlations of the boundary theory are determined by inserting probe fields in the `bulk' gravity theory, and computing
the limiting values of their Green's functions near the boundary. As representative examples of such correlations, we introduce
the simplest example of `matter' fields in the gravity theory: a complex scalar field $\phi$, and a Dirac fermion, $\Psi$.
The action of these fields respectively has the schematic form
\bea
\mathcal{S}_b &=& \int d^4 x \sqrt{-g} \left[ - |\nabla \phi - i A \phi|^2 - m^2 |\phi |^2 \right] \nn
\mathcal{S}_f &=& \int d^4 x \sqrt{-g} \left[ - \overline{\Psi} \Gamma \cdot \left( \partial + \frac{1}{4} \omega_{ab} \Gamma^{ab} - i A \right) \Psi 
- m \overline{\Psi} \Psi \right] \label{Sm}
\eea
where $\omega_{ab}$ is the spin connection of the curved geometry, and the $\Gamma$ represent Dirac matrices.
The parameter $m$ determines the scaling dimension of the operator being used to probe the compressible quantum states.
We are now faced with the problem of determining the inverses of the differential operators in Eq.~(\ref{Sm}) in the 4-dimensional
space defined by Eqs.~(\ref{bh}) and (\ref{fu3}), and taking their limit $u \rightarrow \infty$ limit. This is an intricate problem
in solving differential equations on a curved space, addressed in recent work \cite{sslee0,hong0,zaanen1,hong1,denef}. In the low frequency ($\omega$) limit, the Green's functions
of the boundary compressible quantum theory had the following form for both the boson and fermion cases \cite{hong1}:
\beq
G^{-1} (k, \omega) = C(k) + D(k) \omega^{\nu_k} \quad; \quad T=0, \label{local}
\eeq
where $C(k)$, $D(k)$, and $\nu_k$ are smooth functions of the wavevector $k$. This peculiar behavior, with a smooth dependence on $k$,
and a singular dependence on $\omega$, is a direct consequence of the AdS$_2 \times R^2$ factorization of the geometry in Eq.~(\ref{ads2}):
the independent quantum impurities have temporal correlations which decay with a power-law in time, but are spatially uncorrelated.
The behavior in Eq.~(\ref{local}) does not correspond to any of the generic compressible condensed matter phases considered
in Section~\ref{sec:cond2}.
Instead, it is essentially the behavior observed in the compressible critical states 
in the limit of large spatial dimension or lattice co-ordination number\cite{sy,pg,pgs,bgg,si1,si2}.
The correspondence between these large dimension models and the holographic solutions extends also to 
the extension of Eq.~(\ref{local}) to non-zero temperatures.

The computation of the inverses of the operators in Eq.~(\ref{Sm}), also yields stability conditions on the
gravity theory. For the boson action, $\mathcal{S}_b$, the condition is $C(k) > 0$. This is
satisfied for $m$ sufficiently large. However, with decreasing $m$ there is instability towards condensation
of the scalar, leading to a new compressible phase with superfluidity \cite{gubser,hhh,sonner,gubserpufurocha,Donos,gauntlett}; 
we will discuss this phase further in 
Section~\ref{sec:beyond}.

For the fermion case, $C(k)$ is allowed to be either positive or negative. If $C(k)$ changes sign as a function of the $k$,
we obtain a Fermi surface \cite{hong1} at $k=k_F$, where $C(k_F)=0$, by Eq.~(\ref{fs}). The singularity of the Green's function 
in Eq.~(\ref{local}) at the Fermi surface is a special case of the generic form in Eq.~(\ref{fsnfl}); the latter is singular as a function
of both $\omega$ and $k-k_F$, and reduces to the singular part of the present form obtained for the AdS$_2 \times R^2$ theory only in the
limit where the dynamic critical exponent $z \rightarrow \infty$. Indeed, the value $z=\infty$ is another way of characterizing
the peculiar phase of the holographic theory with a geometry with factorization between space and time.

Finally, we should compare the compressible state described by the Reissner-Nordstr\"om geometry 
to the conjecture towards the end of Section~\ref{sec:cond2}. We have obtained a compressible state which
preserves both the translational and global U(1) symmetry associated with $\mathcal{Q}$. 
We found Fermi surfaces only over a limited parameter regime, but even when present, they enclose
areas which do not obey the Luttinger relation in Eq.~(\ref{lut}). For the conjecture to be valid,
the deficit in the Luttinger relation
must be made up by `hidden' Fermi surfaces associated with fermions which carry additional gauge charges,
as in the non-Fermi liquid and fractionalized Fermi liquid states discussed in Section~\ref{sec:cond2}. The correlators
of fermions with gauge charges are not readily computable in the gauge gravity duality, and are a worthy topic 
for future research.

\subsubsection{Beyond AdS$_2 \times R^2$}
\label{sec:beyond}

Given the non-zero ground state entropy density of the compressible state with $z=\infty$ theory described in Section~\ref{sec:holo2}, 
it seems clear we need to move beyond the Reissner-Nordstr\"om geometry obtained via the equations
of motion of the Einstein-Maxwell action in Eq.~(\ref{Sem}). In particular, at the very least, we should include the 
back-reaction of the matter fields in Eq.~(\ref{Sm}) upon the geometry.
This is an active topic of current research, and I summarize a few
recent developments.

\begin{figure}[htbp]
  \centering
  \includegraphics[width=4.4in]{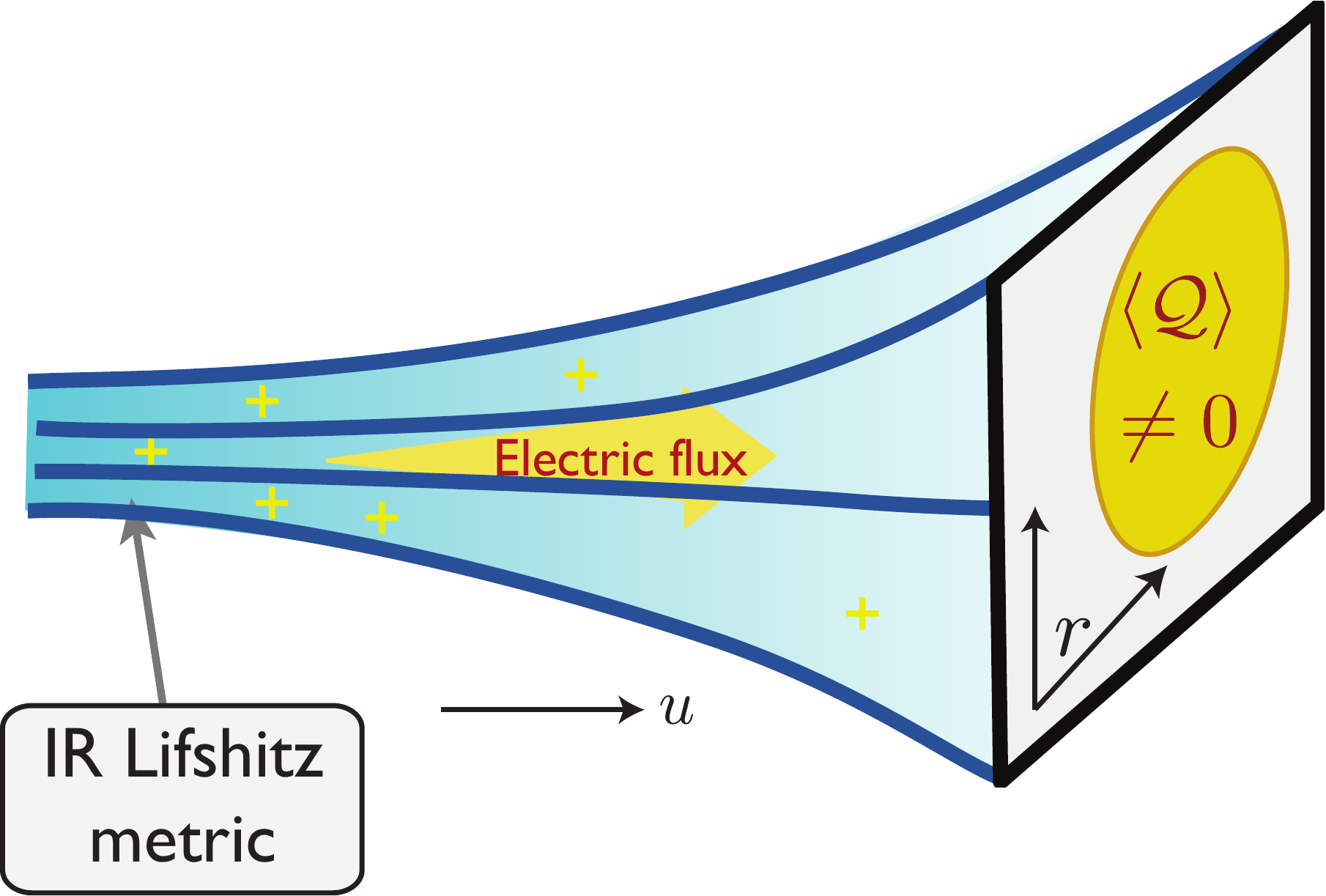}
  \caption{Schematic of the geometry for compressible quantum matter, after accounting for back-reaction
  from the bulk matter in Eq.~(\ref{Sm}) in a Thomas-Fermi or Bose-Einstein theory. There is no horizon, and the small $u$ geometry has a `Lifshitz' 
  form, as in Eq.~(\ref{lif}). The bulk charge density screens the electric field, so that the electric field vanishes as $u \rightarrow 0$.}
  \label{lifshitz}
\end{figure}

As an example of a back-reaction on the metric, we can move to the superfluid state where the bosons are condensed,
and recompute \cite{nellore,roberts} 
the metric $g$ and gauge field $A_t$ from the classical equations of motion of $\mathcal{S}_{EM} + \mathcal{S}_b$,
where $\phi$ is treated as a c-number as in the Bose-Einstein theory at $T=0$. 
In the non-superfluid state, some studies \cite{polchinski,eric,sean1,larus2,pp} have accounted for the back reaction
of fermions in the Thomas-Fermi approximation, where the fermion density at a given $u$ is a function only of
the local chemical potential. In both of these cases, a qualitatively similar back-reaction is obtained, which is illustrated in Fig.~\ref{lifshitz}.
The electric field is screened by the bosonic or fermionic matter in the bulk, so that the electric field vanishes as $u \rightarrow 0$.
In the opposite limit $u \rightarrow \infty$, the electric field remains pinned by the value of $\langle \mathcal{Q} \rangle$, 
as is required by Gauss' Law on the boundary. For the metric, the following qualitative structure is obtained as $u \rightarrow 0$
\beq
ds^2 =  \alpha \frac{du^2}{u^2} - u^{2z} dt^2 + u^2 dr^2 ,
\label{lif}
\eeq
where $\alpha$ and $z$ are constants.
It has become the practice in the string theory literature to call this the `Lifshitz' metric \cite{lifshitz} (for not particularly good reasons).
A key property of this metric is that it is invariant under the following transformations, which generalize (\ref{scal}) and (\ref{scalu}),
\beq
t \rightarrow t/b^z \quad,\quad r \rightarrow r/b \quad, \quad u \rightarrow b \, u ,
\label{scalz}
\eeq
and identify $z$ as the dynamic critical exponent. The AdS$_4$ metric of Eq.~(\ref{ads}) corresponds to $z=1$.
The AdS$_2 \times R^2$ metric of Eq.~(\ref{ads2}) is obtained in the limit $z \rightarrow \infty$ (as expected),
after setting $u = \tilde{u}^{1/z}$.

The existence of this scale-invariant structure in the small $u$ limit is puzzling, and is not well understood.
It implies there are emergent low energy excitations which
scale with the dynamic critical exponent $z$. In the traditional condensed matter superfluid state, the only low energy
excitations of a superfluid are the Goldstone modes, and there are no dynamic critical fluctuations. 
This indicates that the holographic superfluids \cite{gubser,hhh} are not conventional superfluids at $T=0$, and perhaps co-exist
with a non-Fermi liquid sector.
For the non-superfluid compressible state obtained using the Thomas-Fermi theory, it is possible that these critical
excitations are associated with the gapless gauge excitations which led to Eq.~(\ref{fsnfl}) in the non-Fermi liquid.
However, the latter excitations are associated with long wavelength, low energy fluctuations at momenta near the Fermi surface,
and these are not contained in the Thomas-Fermi theory. Indications from recent work \cite{leiden,trivedi,stellar,hong4,ssfl}
are that these low-energy excitations arise from the presence of 
an {\em infinite\/} number of Fermi surfaces, with a Fermi wavevector which varies continuously with $u$.

\begin{figure}[htbp]
  \centering
  \includegraphics[width=4.4in]{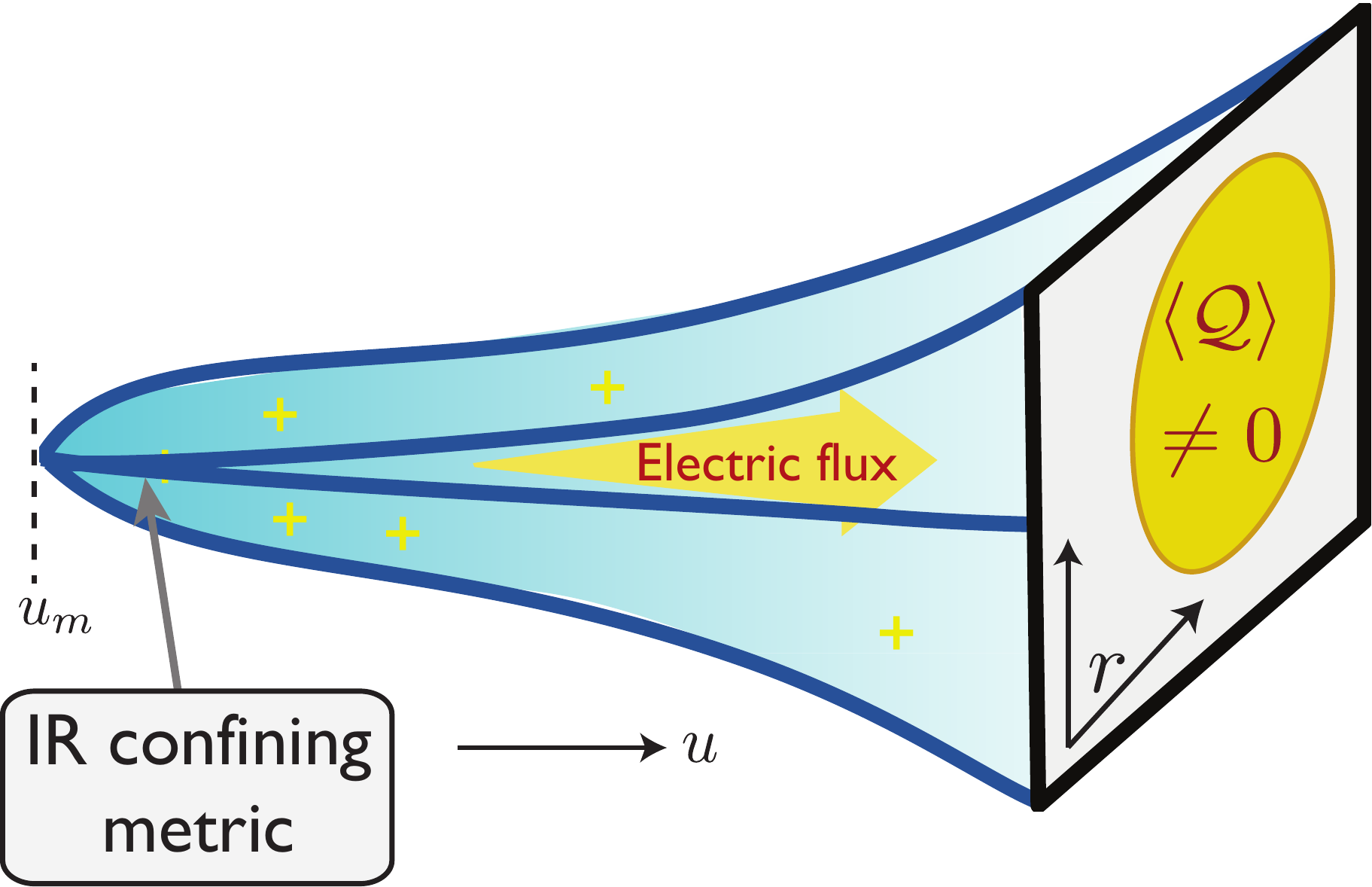}
  \caption{Confining geometry for a Fermi liquid state. The geometry terminates at $u=u_m$ and the
  bulk charge density screens the electric field, so that the electric field vanishes as $u \rightarrow u_m$.}
  \label{confine}
\end{figure}

For condensed matter applications, we therefore need to move the theory of compressible states to a regime with a small number
of Fermi surfaces. This was addressed recently in Ref.~\onlinecite{ssfl} by filling up fermionic states of $\mathcal{S}_f$
in a suitable metric. A key point made in  Ref.~\onlinecite{ssfl} is that a Fermi liquid must be a confining state of the 
gauge theory defining the CFT, as we discuss in the Supplementary Material. In holographic studies of zero density theories of particle physics,
it is known that a confining state corresponds to a geometry which terminates at finite $u=u_m$, rather than extending
all the way to $u=0$; see Fig.~\ref{confine}. 
Assuming such a geometry, Ref.~\onlinecite{ssfl} discussed a complete quantum solution of the theory $\mathcal{S}_f$
in Eq.~(\ref{Sm}), without making the Thomas-Fermi approximation. A conventional Fermi liquid state was found, without extraneous
excitations. Moreover, the Luttinger relation in Eq.~(\ref{lut}) was found to be exactly obeyed with a small number of Fermi surfaces,
including the simplest case with a single Fermi surface; this Luttinger relation was a consequence of Gauss's Law for the electric flux
in the $u$ direction. Confining geometries were also considered for the superfluid state \cite{tadashi,gary}, and
our present arguments indicate that these are traditional superfluids.

Holographic realizations of the familiar compressible phases of condensed matter physics, Fermi liquids and superfluids,
thus finally appear to have been found. Extensions of this understanding to non-Fermi liquids and fractionalized Fermi liquids are important topics for future research. The connection between the Luttinger relation and Gauss' law \cite{ssfl} indicates that the holographic realizations of these phases will have at least part of the electric flux `leaking out' at $u=0$, as in Fig.~\ref{rnbh}, and in a recent study \cite{trivedi}.

\section{Conclusions}
\label{sec:conc}

We divided our discussion into two classes of systems realizable in condensed matter physics: conformal and compressible.

Conformal systems were discussed in Section~\ref{sec:cft}. These concerned generic models which realize conformal
field theories (CFTs) at a quantum phase transition, the simplest being the superfluid-insulator transition of bosons at integer
filling in a periodic potential in two spatial dimensions. Many $T=0$ properties of the CFT are accurately computable
by traditional field theoretic perturbative expansions. However, this success does not extend to long-time correlations
at $T>0$, because the orders of limit of $t \rightarrow \infty$ and $T \rightarrow 0$ do not commute for
all such methods. The Boltzmann equation, and its many descendants, are traditionally used to resum the 
perturbative methods, and access the long time limit; these are only expected to be reasonable when particle-like
excitations are long-lived. The holographic method provides an alternative description of relaxational
and transport properties at long time: it does not assume any picture of particle-like excitations and so is far removed from,
and complementary to, the Boltzmann description. 
We described a gradient expansion method \cite{ajay} for generating an effective holographic theory for generic CFTs.
The coupling constants of this effective theory can be fixed by matching to various $T=0$ correlators of the CFT.
Then the effective theory generates useful results for the long-time limit, including the values of transport co-efficients.
I believe this method offers hope for quantitative, testable predictions not only for the linear transport problems
discussed in Section~\ref{sec:holo1}, but also non-linear and non-equilibrium \cite{sondhi} problems in the vicinity
of quantum critical points.

Compressible systems were studied in Section~\ref{sec:cond2}. The familiar condensed matter examples
are superfluids, solids, and Fermi liquids (and their variants). An exotic class of compressible
states have emerged in condensed matter studies motivated by the strange metal problem of correlated
electron systems \cite{physicstoday}. These are varieties of non-Fermi liquids, all of which have
sharp Fermi surfaces, but the fermionic quasiparticles near some or all of the Fermi surfaces are not well-defined because of their
strong coupling to deconfined gauge fields, or to gapless modes at a quantum phase transition.
This led us to the conjecture \cite{liza}, at the end of Section~\ref{sec:cond2}, that all compressible, continuum
quantum systems which do not break translational or the global U(1) symmetry (associated 
with the conserved density) at $T=0$, must have Fermi surfaces; further, the areas enclosed
by these Fermi surfaces must obey the Luttinger relation in Eq.~(\ref{lut}). 

The holographic method has the potential to provide us with an alternative classification, and a deeper understanding 
of the compressible
states of quantum matter. Such a classification should surely include the familiar solids, superfluids, and Fermi liquids,
and it is not yet settled if the exotic compressible states obtained via holography correspond to those obtained so far
in the condensed matter literature. It remains possible that the Fermi surface conjecture is false, and holography leads us to qualitatively
new types of compressible phases.
 
The Reissner-Nordstr\"om solution of the Einstein-Maxwell action
provides the simplest theory of a compressible state. We reviewed the physical properties of this state:
a non-zero ground state entropy, `locally critical' correlations in time, a non-Fermi liquid Fermi surface with dynamic critical exponent 
$z=\infty$, and 
$T>0$ correlations similar to those of critical quantum impurity problems.
We argued that all these properties were in close correspondence with the `fractionalized Fermi liquid' phase \cite{ffl1,ffl2} of lattice quantum
models solved in the limit of large spatial dimensionality or large lattice co-ordination number \cite{ssffl}.
In Section~\ref{sec:beyond} we discussed ongoing research moving beyond the Reissner-Nordstr\"om geometry.
In many studies, back-reaction of the matter fields on the gravity theory was examined in Thomas-Fermi-like
approximations, and led to a low energy
`Lifshitz' geometry in Eq.~(\ref{lifshitz}) with a finite $z$. However, the physical interpretation of
the critical low energy excitations with such a geometry remains unclear: such excitations
are absent from conventional superfluids and Fermi liquids, and it is unlikely they correspond
to the modes of known non-Fermi liquids. They may well be artifacts of the presence of an infinite number of Fermi surfaces,
with a continually varying Fermi wavevector, along the emergent holographic direction \cite{ssfl}.
In other studies \cite{tadashi,gary,ssfl}, confining geometries
which terminate along the holographic $u$ direction have been considered, and these do realize
traditional superfluids and Fermi liquids, without extraneous excitations. These studies also indicate that the compressible non-Fermi liquid (or fractionalized Fermi liquid) phases will be realized in geometries with 
all (or part) of the electric flux remaining unscreened until the end of the holographic direction.

It does seem that further refinements of the theory are needed
before the links between the condensed matter and holographic theories of compressible
quantum matter are completely established.
Given the pace of progress in the past few years, we can hope that many of these issues
will be resolved in the not too distant future.\\
{\em Note added:} A new approach to holographic theories of compressible quantum matter
has since appeared in Ref.~\onlinecite{hyper}.

\subsection*{Acknowledgements}

I thank Liza Huijse for valuable comments on the manucript.
This research was supported by the National Science Foundation under grant DMR-1103860 and by a MURI grant from AFOSR.

\newpage
\setcounter{section}{0}
\setcounter{figure}{0}
~\\~\\
\begin{center}
{\large\bf Supplementary Material:\\ Conformal and Compressible Quantum Matter}\\
~\\
Subir Sachdev\\
{\em Department of Physics, Harvard University, Cambridge MA
02138}\\
~\\~\\
{\large Abstract}\\
Field theories
of conformal and compressible states of matter\\ are derived from lattice Hubbard models.
\end{center}
\vspace{0.5in}
\setcounter{page}{1}

These notes are adapted from earlier reviews \cite{bbssp,tasip}.

\section{Conformal quantum matter}
\label{sec:conf}

We will demonstrate that the superfluid-insulator quantum phase transition 
of the boson Hubbard model in two spatial dimensions is described by a conformal field theory (CFT).

The boson Hubbard model is (as in Eq.~(\ref{hubbard}) in the main paper)
\beq
H_b = - w \sum_{\langle ij \rangle} \left( b^\dagger_i b_j + b_j^\dagger b_i \right) + \frac{U}{2} \sum_i n_i (n_i - 1)
- \mu \sum_i n_i ,
\label{hubbardp}
\eeq
where $b_i$ is the canonical boson annihilation operator, 
$n_i = b^\dagger_i b_i$ is the boson number operator, $w$ is the hopping matrix element between nearest-neighbor
sites, $U$ is the on-site repulsive energy between a pair of bosons, and $\mu$ is the chemical potential. 
Let us assume that the average boson density is exactly $n_0$
per site, where $n_0$ is a positive integer.  For $U/w \gg 1$, the ground
state is simply
\begin{equation}
  \label{eq:MGS}
  |GS\rangle = \prod_i (b_i^\dagger)^{n_0} |0\rangle,
\end{equation}
where $|0\rangle$ is the empy state with no bosons.  
In the same limit, the lowest
excited states are ``particles'' and ``holes'' with one extra or
missing boson,
\begin{eqnarray}
  \label{eq:Mex}
  |p_i\rangle & = & b_i^\dagger |GS\rangle, \\
  |h_i\rangle & = & b_i |GS\rangle.
\end{eqnarray}
For $w/U=0$ strictly, the particle and hole energies (relative to the
ground state) are
\begin{equation}
  \label{eq:Eph}
  E^{(0)}_p = U n_0 - \mu , \qquad E^{(0)}_h = U(1 - n_0) + \mu. 
\end{equation}
For $0<w/U \ll 1$, these states will develop dispersion.
By
considering the first order splitting of the degenerate manifold of
particle or hole states (degeneracy associated with the site of the
particle or hole), one obtains (considering the square lattice for simplicity)
\bea
  \label{eq:Epht}
  E_p ({\bm k}) &=& E^{(0)}_p - 2 w (n_0+1)(\cos k_x + \cos k_y)\approx
  \Delta_{p} + \frac{k^2}{2m_p} \nn
  E_h ({\bm k}) &=& E^{(0)}_h - 2 w n_0 (\cos k_x + \cos k_y)\approx
  \Delta_{h} + \frac{k^2}{2m_h},
\eea
where we have Taylor expanded around the minimum at ${ k=0}$,
giving $m_p = 1/(2(n_0+1)w)$ and  $m_h = 1/(2 n_0 w)$ 
The excitation gaps, $\Delta_{p/h}$, are
\bea
  \label{eq:Egap}
  \Delta_{p} &=& U n_0 - \mu - 4 w(n_0 + 1) \nn
  \Delta_{h} &=& U (1- n_0) + \mu - 4 w n_0 .
\eea
to this order in $w/U$. As long as both of the these gaps are positive, our starting point of a 
Mott insulating state with an average of $n_0$ particles per site is stable.

When one of the gaps vanish, the Mott insulator is no longer stable,
and we have a quantum transition to a superfluid state. Let us assume that it is $\Delta_p$ that vanishes
first with increasing $w$. The transition then corresponds to a Bose-Einstein condensation of particles, with $-\Delta_p$ 
acting as the effective chemical potential. At $T=0$, an increasing chemical potential implies an increasing particle
density, and so the superfluid state will have a density greater than that of the Mott insulator.
Similarly, if the value of $\mu$ is such that $\Delta_h$ vanishes first, the superfluid state will have a density smaller than
that of the Mott insulator.

However, let us consider the special case when the density of both the Mott insulator and the superfluid are equal to $n_0$;
this is often naturally the case under experimental conditions. Our reasoning makes it clear that this is only possible
if $\mu$ is chosen so that $\Delta_p=\Delta_h \equiv \Delta$.
This both gaps vanish simultaneously, the insulator-superfluid transition corresponds
to condensation of both particles and holes (which can be viewed as ``anti-particles'').
This symmetry between particles and anti-particles is responsible for the relativistic structure
of the low energy theory.

Let us now proceed to derive the effective action for the low energy
theory near the insulator-superfluid transition.
While it is possible to derive a field theory of this condensation
from $H_b$, we instead just write it down based on our simple
physical picture.  We model the particle and hole excitations by
fields $p(r,\tau), h(r,\tau)$ respectively, in the
imaginary time ($\tau$) path integral.  The weight in the path
integral is, as usual, the Euclidean action,
\begin{equation}
  \label{eq:Sph}
   \mathcal{S}_b = \int\!d\tau d^2 r\, \left[ p^\dagger \left(\frac{\partial}{\partial \tau} +
  \Delta - \frac{\nabla^2}{2m_p}\right)p^{\vphantom\dagger} + h^\dagger\left(
  \frac{\partial}{\partial \tau} +
  \Delta - \frac{\nabla^2}{2m_h}\right)h^{\vphantom\dagger} - \Lambda(p^\dagger
  h^\dagger + p^{\vphantom\dagger}h^{\vphantom\dagger}) + \cdots\right].
\end{equation}
Here we have included a term $\Lambda$ which creates and annihilates
particles and holes together in pairs, which is expected since this
conserves boson number.  Microscopically this term arises from the
action of the hopping $w$ on the naive ground state, which creates
particle-hole pairs on neighboring sites, so $\Lambda \sim O(w)$ (the
spatial dependence is unimportant for the states near ${ k=0}$).
We have neglected -- for brevity of presentation -- to write a number
of higher order terms involving four or more boson fields,
representing interactions between particles and/or holes, and other
boson number-conserving two-body and higher-body collisional
processes.  Note that the dependence upon $w/U$ in Eq.~(\ref{eq:Sph})
arises primarily through implicit dependence of $\Delta$.

Without loss of generality, we assume $\Lambda > 0$, and change 
 variables to the linear combinations
\begin{equation}
  \label{eq:lc}
  \psi = \frac{1}{\sqrt{2}}(p^{\vphantom\dagger}+h^\dagger) \qquad
  \xi = \frac{1}{\sqrt{2}}(p^{\vphantom\dagger}-h^\dagger). 
\end{equation}
Then the quadratic terms in the action are
\bea
\mathcal{S}_b &=& \int d \tau d^2 r\, \Biggl[ \xi^\dagger \frac{\partial\psi}{\partial \tau} - \xi 
\frac{\partial\psi^{\dagger}}{\partial \tau} 
+ (\Delta - \Lambda) |\psi|^2 + (\Delta + \Lambda) |\xi|^2 \nn 
&+& \left( \frac{1}{4 m_p} + \frac{1}{4 m_h} \right) (|\nabla \psi|^2 + |\nabla \xi|^2) 
+ \left( \frac{1}{4 m_p} - \frac{1}{4 m_h} \right) (\nabla \psi^{\dagger} \nabla \xi +
\nabla \xi^{\dagger} \nabla \psi ) 
\Biggr].
\eea
Notice that the quadratic form for $\psi$ becomes
unstable, before that of $\xi$. So let us integrate out $\xi$, expanding the resulting action
in powers and gradients of $\psi$.
In this manner we obtain the theory for the superfluid order parameter $\psi$ \cite{FWGFp}
\bea
\mathcal{S}_b &=& \int d \tau d^2 r\, \left[ \frac{1}{(\Delta + \Lambda)} \left|\frac{\partial\psi}{\partial \tau} \right|^2 + (\Delta - \Lambda) |\psi|^2  
+ \left( \frac{1}{4 m_p} + \frac{1}{4 m_h} \right) |\nabla \psi|^2 + u |\psi|^4 \right].
\label{lgw}
\eea
This is the promised relativistic field theory, written in Eq.~(\ref{sb}) in the main paper. 
The energy gap for both particle and hole excitations
is $\sqrt{\Delta^2 - \Lambda^2}$, and this vanishes at the quantum critical point.

There are a number of additional higher-order terms, not displayed above, which are not relativistically
invariant. However, all of these are formally irrelevant at the Wilson-Fisher fixed point which controls
the critical theory. 

Finally, we note that the scaling properties and relativistic invariance of the critical point are sufficient to 
establish its invariance under conformal transformations.

\subsection{Quantum critical transport}
\label{sec:qctrans}

To illustrate the general issues, we begin by computing the transport properties of the
free field theory of a complex scalar with mass $m$, written in a Lorentz invariant notation:
\beq
\mathcal{S}_\psi = \int d^D r \left[ |\partial_\mu \psi|^2 + m^2 |\psi |^2 \right]
\label{free}
\eeq
This theory can be obtained from Eq.~(\ref{lgw}) at $u=0$, after appropriate rescalings of co-ordinates
and fields.

The conserved electrical current is
\beq J_\mu = -i \left( \psi^\ast \partial_\mu \psi - \psi \partial_\mu \psi^\ast \right). \label{Jmu}
\eeq
Let us compute its two-point correlator, $K_{\mu\nu} (k)$ at a spacetime momentum $k_\mu$ at $T=0$.
This is given by the one-loop diagram which evaluates to
\bea K_{\mu\nu} (k) &=& \int \frac{d^3 p}{8 \pi^3} \left( \frac{ (2p_\mu + k_\nu)(2 p_\nu + k_\nu) }{((p+k)^2 + m^2)(p^2 + m^2)}
- 2 \frac{\delta_{\mu\nu}}{p^2 + m^2} \right)
 \nn
&=&- \frac{1}{8\pi} \left( \delta_{\mu\nu} - \frac{k_\mu k_\nu}{k^2} \right) 
\int_0^1 dx \frac{k^2 (1- 2x)^2}{\sqrt{m^2 + k^2 x (1-x)}}.
\label{Kmn}
\eea
The second term in the first equation arises from a  `tadpole' contribution which is omitted in Eq.~(\ref{Jmu}). 
Note that the current correlation is purely transverse, and this follows from the requirement
of current conservation
\beq k_\mu K_{\mu\nu} = 0. \label{ccons} \eeq
Of particular interest to us is the $K_{00}$ component, after analytic continuation to
Minkowski space where the spacetime momentum $k_\mu$ is replaced by $(\omega, k)$.
The conductivity is obtained from this correlator via the Kubo formula
\beq
\sigma (\omega) = \lim_{k \rightarrow 0} \frac{-i \omega}{k^2} K_{00} (\omega, k). \label{kubo}
\eeq

In the insulator, where $m > 0$, analysis of the integrand in Eq.~(\ref{Kmn}) shows that that the spectral weight of the 
density correlator has a gap of $2m$ at $k=0$, and the conductivity in Eq.~(\ref{kubo}) vanishes.
These properties are as expected in any insulator.

At the critical point, where $m=0$, the fermionic spectrum is gapless, and so is
that of the charge correlator. The density correlator in Eq.~(\ref{Kmn}) and
the conductivity in Eq.~(\ref{kubo}) evaluate to the simple universal results
\bea
K_{00} (\omega, k) &=& \frac{1}{16} \frac{k^2}{\sqrt{k^2 - \omega^2}} \nn
\sigma (\omega) &=& \frac{1}{16}. \label{univ}
\eea

Going beyond the free field theory in Eq.~(\ref{free}), the effect of interactions can be accounted
for order-by-order in $u$. In the renormalization group approach, $u$ takes the value specified by
the Wilson-Fisher fixed point at the quantum critical point. Combined with the absence of of divergencies in the
perturbative expansion (which is a consequence of Eq.~(\ref{ccons})), this means the only effect
of interactions is to change the pre-factor in Eq.~(\ref{univ}) to a different universal numerical value.
So we write
\bea
K_{00} (\omega, k) &=& \sigma_\infty \frac{k^2}{\sqrt{k^2 - \omega^2}} \nn
\sigma (\omega) &=& \sigma_\infty, \label{univ2}
\eea
where $\sigma_\infty$ is a universal number dependent only upon the universality class of the 
quantum critical point, 
whose value can be computed by various expansion methods.

\subsubsection{Non-zero temperatures}

We begin by repeating the above computation for the free field theory at $T>0$.
This only requires replacing the integral over the loop frequency in Eq.~(\ref{Kmn}), by a summation
over the Matsubara frequencies which are quantized by integer multiples of $2\pi T$.
Such a computation, via Eq.~(\ref{kubo}) leads to the conductivity 
\bea
\mbox{Re}[\sigma (\omega)]  &=& \mathcal{P} \, \delta (\omega) + \frac{\theta(|\omega| - 2m)}{16} \left( \frac{ \omega^2 - 4 m^2}{4 \omega^2} \right) \coth \left( \frac{|\omega|}{4T} \right) \nn
\mathcal{P} & \equiv & \frac{1}{8 T} \int_0^\infty \frac{ k^3 dk}{(k^2 + m^2)} \frac{1}{\sinh^2 \left( {\sqrt{k^2 + m^2}}/{2 T} \right) };
\label{drudep}
\eea
the imaginary part of $\sigma (\omega)$ is the Hilbert transform of $\mbox{Re}[\sigma (\omega)] -1/16$.
Note that this reduces to Eq.~(\ref{univ}) in the limit $\omega \gg T$.
However, the most important new feature of Eq.~(\ref{drudep}) arises for $\omega \ll T$,
where we find a delta function at zero frequency in the real part.
Thus the d.c. conductivity is infinite at this order, arising from the collisionless transport of thermally excited
carriers. This is clearly an artifact of the free field theory.

At non-zero $u$,
collisions between carriers invalidate the form in Eq.~(\ref{drudep}) for the density
correlation function, and we instead expect the form dictated by the hydrodynamic diffusion of charge.
Thus for $K_{00}$, Eq.~(\ref{univ2}) applies only for $\omega \gg T$, while
\beq
K_{00} (\omega, k)  = \chi \frac{D k^2}{Dk^2 - i \omega} \quad, \quad \omega \ll T. \label{diff}
\eeq
Here $\chi$ is the charge susceptibility (here it is the compressibility),
and $D$ is the charge diffusion constant. By the universality of the Wilson-Fisher fixed point, we expect
that these have universal
values in the quantum critical region:
\beq
\chi = \mathcal{C}_\chi T \quad , \quad D = \frac{\mathcal{C}_D}{T},
\eeq
where again $\mathcal{C}_\chi$ and $\mathcal{C}_D$ are universal numbers. 
For the conductivity, we expect a crossover from the collisionless critical dynamics at
frequencies $\omega \gg T$, to a hydrodynamic collision-dominated form for $\omega \ll T$.
This entire crossover is universal, and is described by a universal crossover function
\beq 
\sigma (\omega)  = \Sigma (\omega/T). \label{Kcross}
\eeq
The result in Eq.~(\ref{univ2}) applies for $\omega \gg T$, and so 
\beq
\Sigma (\infty) = \sigma_\infty . \label{Kinf} \eeq
For the hydrodynamic transport, we apply the Kubo formula in Eq.~(\ref{kubo}) to Eq.~(\ref{diff})
and obtain
\beq
\Sigma (0) = \mathcal{C}_\chi \mathcal{C}_D
\eeq
which is a version of Einstein's relation for Brownian motion.

\section{Compressible quantum matter}
\label{sec:tri}

Now we will consider the Hubbard model for fermionic particles with spin $S=1/2$ (electrons) on the 
triangular lattice. 

For small $U/w$, the ground state of this model is a metal, rather than a superfluid.
This is because the fermions cannot condense; instead they occupy all single particle states inside
a `Fermi surface' in momentum space, forming a Fermi liquid. 

For large $U/w$, and with a density
of one electron per site, we do expect an insulating state to form, with a gap to both particle
and hole excitations, just as was the case for bosons. However, the electron localized on each
site of the lattice now has a spin degeneracy, and we also have to specify the spin wavefunction
in the insulator. At the largest values of $U/w$, it is believed that the insulator has long-range
antiferromagnetic order; we will not study this ordered state here. The nature of the insulator at smaller
$U/w$, and in particular, in the vicinity of the insulator-metal transition is still a question of some debate.
In the following, we will assume that the insulating state proximate to the critical point is a 
particular ``U(1) spin liquid'', which we will describe more completely below.

The Hubbard Hamiltonian is
\beq
H = - w \sum_{\langle ij \rangle}  \left( c^{\dagger}_{i \alpha} c_{j \alpha}
+ c^{\dagger}_{j \alpha} c_{i \alpha} \right) + \sum_i \left[- \mu \left( n_{i \uparrow} + n_{i \downarrow} \right) +  U \left(n_{i \uparrow} - \frac{1}{2} \right)
\left(n_{i \downarrow} - \frac{1}{2}\right)\right]. \label{h1}
\eeq
Here $c_{i\alpha}$, $\alpha = \uparrow, \downarrow$ are annihilation operators on the site $i$ of a 
triangular lattice. The density of electrons is controlled by the chemical potential $\mu$ which couples
to the total electron density, with
\beq
n_{i \uparrow} \equiv c_{i \uparrow}^\dagger c_{i \uparrow} \quad , \quad n_{i \downarrow} \equiv c_{i \downarrow}^\dagger c_{i \downarrow}. \label{h2}
\eeq
For completeness, we also note the algebra of the fermion  operators:
\bea
c_{i \alpha} c_{j \beta}^\dagger + c_{j \beta}^\dagger c_{i \alpha} &=& \delta_{ij} \delta_{\alpha\beta} \nn
c_{i \alpha} c_{j \beta} + c_{j \beta} c_{i \alpha} &=& 0. \label{h3} 
\eea

Let us begin by considering the case $U=0$.
Then the ground state is a metal at all densities, with a Fermi surface separating occupied
and empty states in momentum space.
Landau's Fermi liquid (FL) theory describes how the free-electron model of a metal
can be extended to non-zero $U$.
For our purposes, we need
only two basic facts: ({\em i\/}) the fermionic excitations near the Fermi surface are essentially non-interacting electrons, and
({\em ii\/}) the area enclosed by the Fermi surface is equal to the electron density---this is Luttinger's theorem, which we state more
explicitly below.

\begin{figure}
\center\includegraphics[width=4in]{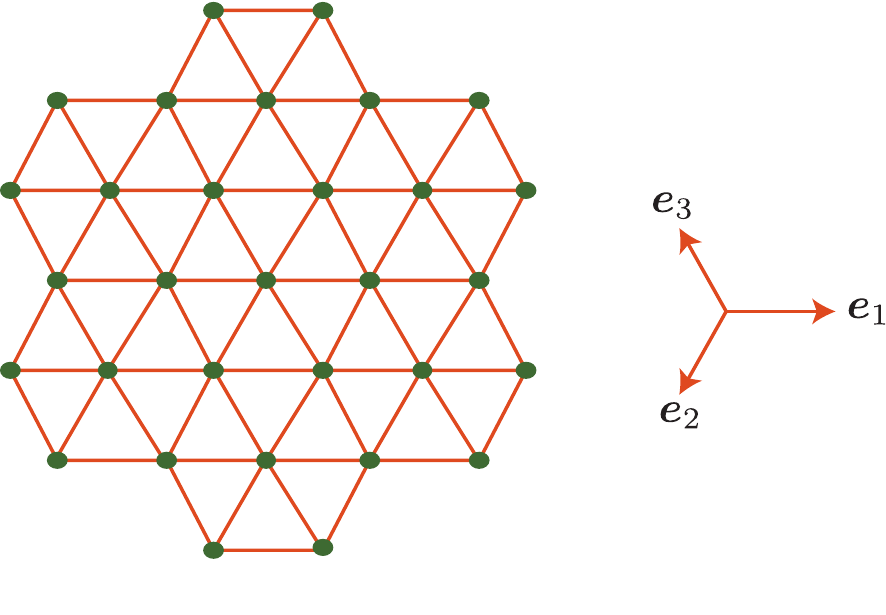}
\caption{The triangular lattice}
\label{fig:trilattice}
\end{figure}

\begin{figure}
\center\includegraphics[width=3in]{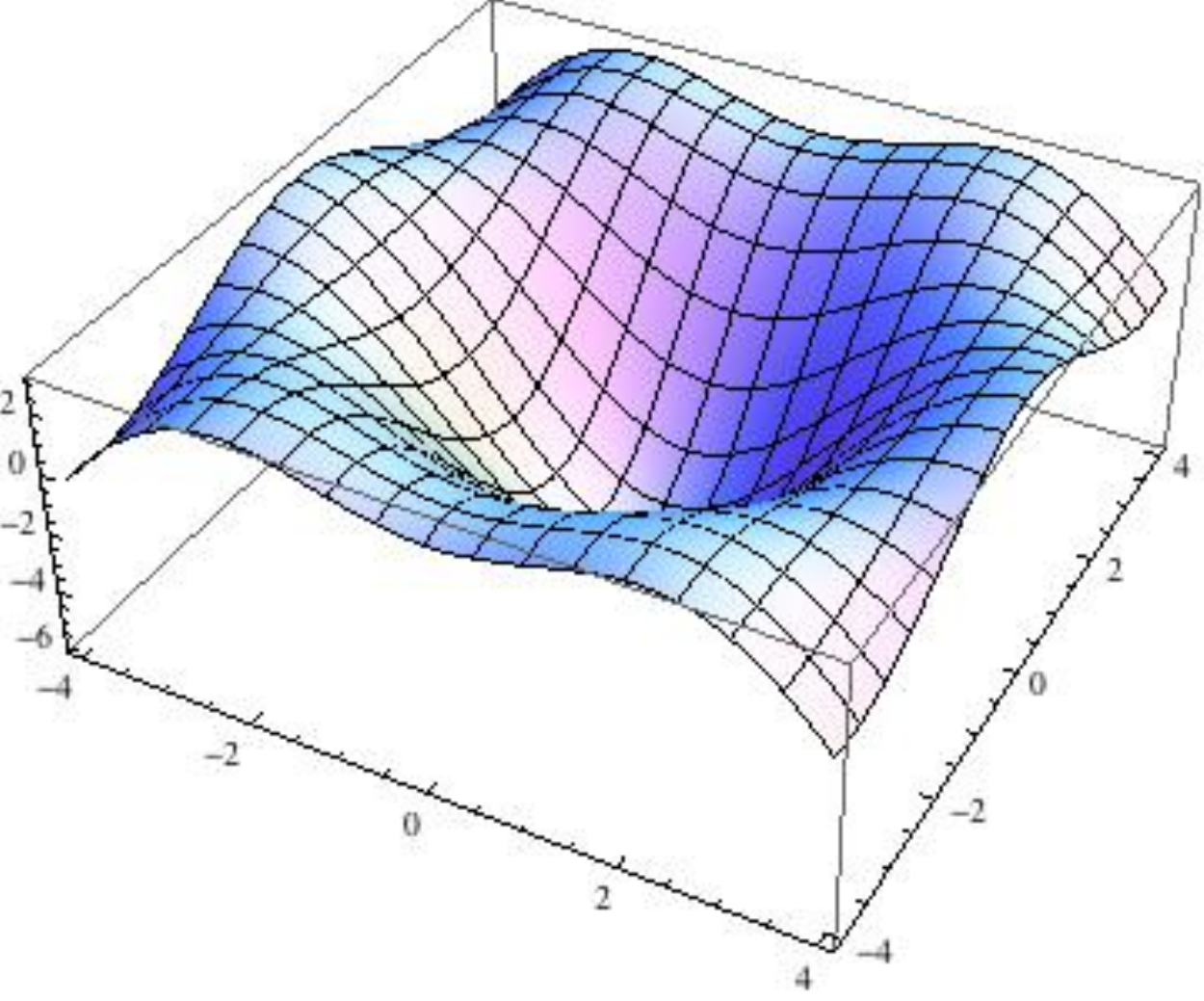}
\caption{The electronic dispersion in Eq.~(\ref{hlattri}) for $\mu=0$ and $t=1$.}
\label{dispt}
\end{figure}

\begin{figure}
\center\includegraphics[width=2.6in]{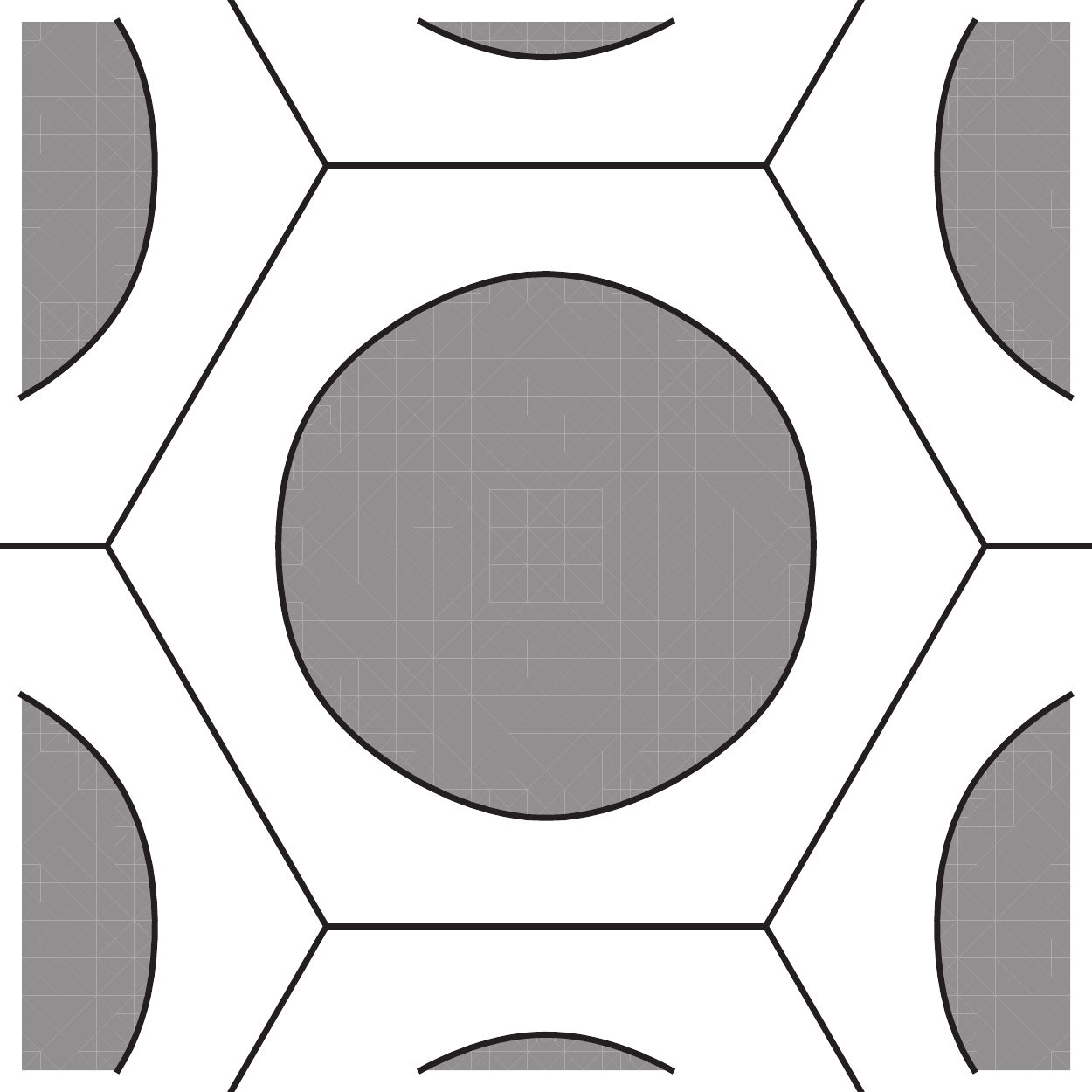}
\caption{The Fermi surface of Eq.~(\ref{hlattri}) for $\mu=1/2$ and $t=1$; the occupied states are shaded. 
Also shown are the periodic images of the Fermi
surface in their respective Brillouin zones.}
\label{fermisurface}
\end{figure}

At $U=0$, the Hamiltonian of the FL metal is
\begin{equation}
H_0 = \sum_{{\bm k}} c_\alpha^{\dagger} ({\bm k}) \Bigl[ - \mu  - 2t \Bigl(\cos({\bm k} \cdot {\bm e}_1) + \cos({\bm k} \cdot {\bm e}_2)
+ \cos({\bm k} \cdot {\bm e}_3) \Bigr) \Bigr] c_\alpha ({\bm k} ), \label{hlattri}
\end{equation}
where the ${\bm e}_i$ are as shown in Fig.~\ref{fig:trilattice}.
The reciprocal lattice consists of the vectors ${\bm G} = \sum_i n_i {\bm G}_i$, where 
\begin{equation}
{\bm G}_1 = \frac{4 \pi}{3} ({\bm e}_1 - {\bm e}_2) \quad , \quad 
{\bm G}_2 = \frac{4 \pi}{3} ({\bm e}_2 - {\bm e}_3) \quad , \quad
{\bm G}_3 = \frac{4 \pi}{3} ({\bm e}_3 - {\bm e}_1). \label{gvecstri}
\end{equation}
The electronic dispersion in Eq.~(\ref{hlattri}) is plotted in Fig.~\ref{dispt}: it only has simple parabolic minima at ${\bm k } = 0$,
and its periodic images at ${\bm k} = {\bm G}$, and there are no Dirac points.
At any chemical potential, the negative energy states are occupied, leading to a Fermi surface bounding the set of occupied states, as
shown in Fig.~\ref{fermisurface}.
Luttinger theorem states that the total area of the occupied states, the shaded region of the first Brillouin zone in Fig.~\ref{fermisurface}
occupies an area, $\mathcal{A}$, given by
\begin{equation}
\frac{\mathcal{A}}{2 \pi^2} = \mathcal{N}, \label{landaufl}
\end{equation}
where $\mathcal{N} = \sum_{\alpha} c^\dagger_{\alpha} c_{\alpha}$ is the total electron density. This relationship is obviously true
for free electrons simply by counting occupied states, but it also remains true for interacting electrons.

Now we turn up the strength of the interactions, $U$, at a density of one electron per site.
By the same argument as that for bosons, an insulator will appear for
sufficiently large $U$.
As stated above, we will focus particular route to the destruction of the small $U$
Fermi liquid, one which reaches directly to an insulator
which is a `spin liquid' \cite{motp,leep,senthilmottp}. 
The spin liquid insulator is a phase in which the spin rotation symmetry is preserved, and there is a
gap to all charged excitations. However, there are gapless spin excitations, and an emergent compact
U(1) gauge field in a deconfined phase.

The key to the description of this metal-insulator transition is an exact rewriting of the Hubbard model in Eq.~(\ref{h1})
as a compact U(1) lattice gauge theory. To derive this, let us proceed with using the same strategy
as that used in Section~\ref{sec:conf} for the boson Hubbard model.
So to represent charged excitations of the insulator, we introduce bosonic operators
$p_i$ and $h_i$, representing the doubly-occupied and empty sites respectively.
However, the singly-occupied site cannot be treated as a featureless vacuum, as we were
able to in Section~\ref{sec:conf}. Now we need a fermionic operator $f_\alpha$ (the `spinon') to
represent the spin orientation of the singly-occupied site. For every site, 
we make the following correpondences for the four
states in the Fock space
\bea 
| 0 \rangle \quad & \Leftrightarrow& \quad h^\dagger | 0 \rangle \nn
c^\dagger_{\alpha} | 0 \rangle \quad & \Leftrightarrow& \quad f^{\dagger}_\alpha | 0 \rangle \nn
c^\dagger_{\uparrow} c^\dagger_{\downarrow} | 0 \rangle \quad & \Leftrightarrow& \quad p^\dagger
f^\dagger_\uparrow f^\dagger_\downarrow | 0 \rangle \label{rotorrep}
\eea
It is now easy to verify that Eq.~(\ref{rotorrep}) is equivalent to the operator identification
\beq
c_\alpha = (h^\dagger + p) f_\alpha , \label{cf}
\eeq
provided we always project to states which obey the constraint
\beq
f_\alpha^\dagger f_\alpha - p^\dagger p + h^\dagger h = 1
\label{fnc} 
\eeq
on every site. All physical observables are operators which stay within the subspace
defined by Eq.~(\ref{fnc}): such operators are invariant under the following
compact U(1) gauge transformation
\beq
f_\alpha \rightarrow f_\alpha e^{i \zeta} \quad , \quad p \rightarrow p e^{-i \zeta} 
 \quad , \quad h \rightarrow h e^{i \zeta}. \label{gauget}
\eeq
As we will show below, there will is an emergent gauge field $B_\mu$ in the effective theory
associated with this gauge transformation. 
The constraint in Eq.~(\ref{fnc}) will be the Gauss law of this gauge theory.
These operator identities are related to those of the `slave rotor' representation \cite{florensp}.

First, let us rewrite the Hubbard model in terms of these new bosonic and fermionic operators.
The Hubbard Hamiltonian in Eq.~(\ref{h1}) is now exactly equivalent to
\bea
H [f,p,h] &=& - w \sum_{\langle ij \rangle} f_{i \alpha}^\dagger f_{j \alpha} (h_i + p_i^\dagger)
(h^\dagger_j + p_j) ~+ ~\mbox{H.c.} \nn
 &~&~~~+ \sum_{i} \left( - \mu (p_i^\dagger p_i - h_i^\dagger h_i + 1) 
 + \frac{U}{2} \left(  p_i^\dagger p_i +  h_i^\dagger h_i - \frac{1}{2} \right) \right), \label{Hfn}
\eea 
provided our attention is restricted to the set of states which obey the constraint in Eq.~(\ref{fnc}) on every lattice site;
note that the Hamiltonian in Eq.~(\ref{Hfn}) commutes with constraints in (\ref{fnc}), and so these can be consistently imposed.
In the on-site terms Eq.~(\ref{Hfn}) we have used the $p$ and $h$ operators  to measure the electron density 
on each site

We can now implement the commutation relations, the Hamiltonian, and the constraint in a coherent state path integral
\bea
\mathcal{Z} &=& \int \mathcal D f_{i \alpha} (\tau)  \mathcal D p_{i} (\tau) 
\mathcal D h_{i} (\tau) \mathcal{D} \lambda_i (\tau) \exp \Biggl( - \int d \tau \, H [f, p, h] \nonumber \\ 
&~&- \int d \tau \sum_i \Biggl[ f^\dagger_{i \alpha} \frac{\partial f_{i \alpha}}{\partial \tau} 
+ p^\dagger_{i} \frac{\partial p_{i}}{\partial \tau} + h^\dagger_{i} \frac{\partial h_{i}}{\partial \tau}
+ i \lambda_i ( f_{i \alpha}^\dagger f_{i \alpha} 
- p_i^\dagger p_i + h_i^\dagger h_i - 1) \Biggr]
 \Biggr),
\label{Ztri}
\eea
where the constraint in Eq.~(\ref{fnc}) is implemented using an auxilliary field $\lambda_i (\tau)$ which acts as a Lagrange multiplier. 

A key observation now is that the partition function in Eq.~(\ref{Ztri}) is invariant under a site, $i$, and $\tau$-dependent
U(1) gauge transformation $\zeta_i (\tau)$ where the fields transform as in Eq.~(\ref{gauget}), and $\lambda$ transforms as
\beq
\lambda_i \rightarrow \lambda_i - \frac{\partial \zeta_i}{\partial \tau}. \label{Lgauge}
\eeq
In other words, $\lambda$ transforms like the temporal component of a U(1) gauge field. 

How do we obtain the spatial components of the gauge field? For this, we apply a ``Hubbard-Stratonovich transformation'' 
to the hopping term in Eq.~(\ref{Hfn}). For this, we introduce another auxiliary complex field $Q_{ij} (\tau)$ which lives on the links of the triangular
lattice and replace the hopping term by
\beq \sum_{\langle ij \rangle} 
\left( \frac{|Q_{ij} (\tau) |^2}{w} - Q_{ij} (\tau) f_{i \alpha}^\dagger f_{j \alpha} - Q_{ij}^\ast (\tau) (h_i + p_i^\dagger)
(h^\dagger_j + p_j)  + \mbox{H.c.} \right)
\eeq
We now see from Eq.~(\ref{gauget}), that $Q_{ij}$ transforms under the gauge transformation in Eq.~(\ref{gauget}) as
\beq
Q_{ij} \rightarrow Q_{ij} e^{i (\zeta_i - \zeta_j)}. \label{Qgauge}
\eeq
In other words, arg($Q_{ij}$) is the needed spatial component of the compact U(1) gauge field.

So far, we have apparently only succeeded in making our analysis of the Hubbard model in Eq.~(\ref{h1}) more complicated. Instead of the 
functional integral of the single complex fermion $c_{i \alpha}$, we now have a functional integral over the complex fermions $f_{i \alpha}$,
the bosons $p_i$, $h_i$, and the auxilliary fields $\lambda_i$ and $Q_{ij}$. How can this be helpful? The point, of course, is that the new
variables help us access new phases and critical points which were inaccessible using the electron operators, and these phases have strong correlations
which are far removed from those of weakly interacting electrons. 

The utility of the new representation is predicated on the assumption that the fluctuations in the auxiliary fields $Q_{ij}$ and $\lambda_i$  
are small along certain directions in parameter space. So let us proceed with this assumption, and describe the structure of the phases so obtained. 
We parameterize
\beq Q_{ij} = \overline{Q}_{ij} e^{B_{ij}} \quad, \quad \lambda_i = - i \overline{\lambda} - B_{i\tau}
\eeq
and ignore fluctuations in the complex numbers $\overline{Q}_{ij}$, and the real number $\overline{\lambda}$. With these definitions, it is clear from Eqs.~(\ref{Lgauge}) and (\ref{Qgauge}) that $B_{ij}$ and $B_\tau$ form the spatial and temporal components of a U(1) gauge field, and so must enter into all physical quantities in a gauge invariant manner. 
The values of $\overline{Q}_{ij}$ and $\overline{\lambda}$ are determined by a suitable saddle-point analysis of the partition function,
and ensure that the constraint (\ref{fnc}) is obeyed.
With these assumptions, the partition function separates into separate fermionic and bosonic degrees of freedom interacting
via their coupling to a common U(1) gauge field $(B_{i\tau}, B_{ij})$. In the continuum limit, the gauge fields become a conventional
U(1) gauge field $B_\mu = (B_\tau, {\bm B})$. The partition function of the gauge theory is
\bea
\mathcal{Z} &=& \int \mathcal D f_{i \alpha} (\tau)  \mathcal{D} p_i (\tau) \mathcal{D} h_i (\tau) \mathcal{D} B_{i \tau} (\tau) \mathcal{D} B_{ij} (\tau) \nonumber \\
&~&~~~~~~~~~~~~~~~~~~ \exp \left( - \int d\tau \Bigl[ \mathcal{L}_f + \mathcal{L}_b
+ i \sum_i B_{i\tau}\Bigr] \right) \nonumber \\
&~&~~~~\mathcal{L}_f =
\sum_i  f^\dagger_{i \alpha} \left( \frac{\partial }{\partial \tau} + \overline{\lambda} - i B_{i \tau} \right)
f_{i \alpha} - \sum_{\langle ij \rangle} \overline{Q}_{ij} f_{i \alpha}^\dagger e^{i B_{ij}} f_{j \alpha} 
+ \mbox{H.c.}
\nonumber \\ 
&~&~~~~\mathcal{L}_b = \sum_i  p_{i}^\dagger \left( \frac{\partial }{\partial \tau} - \overline{\lambda} - \mu
+ \frac{U}{2} + i B_{i \tau} \right)p_i + \sum_i  h_{i}^\dagger \left( \frac{\partial }{\partial \tau} + \overline{\lambda} + \mu + \frac{U}{2} - i B_{i \tau} \right)h_i \nonumber \\
&~&~~~~~~~~~~~~
- \sum_{\langle ij \rangle } \overline{Q}_{ij}^\ast 
 (h_i + p_i^\dagger) e^{-i  B_{ij}}
(h^\dagger_j + p_j) 
 + \mbox{H.c.}.
\label{rotorfermiongauge}
\eea
Thus we have fermions $f_{i\alpha}$ moving in a band structure which is roughly the same as that of the electrons in 
Eq.~(\ref{hlattri}), the bosons $p_i$, $h_i$ Hubbard-like Hamiltonian essentially identical in 
form to that in Section~\ref{sec:conf}, and all particles are minimally coupled to a
compact U(1) gauge field.

We begin by neglecting the gauge fields, and computing the separate phase diagrams of $\mathcal{L}_f$ and $\mathcal{L}_b$.

The fermions are free, and so occupy the negative energy states determined by the chemical potential $\overline{\lambda}$.

\begin{figure}
\centering
 \includegraphics[width=4.2in]{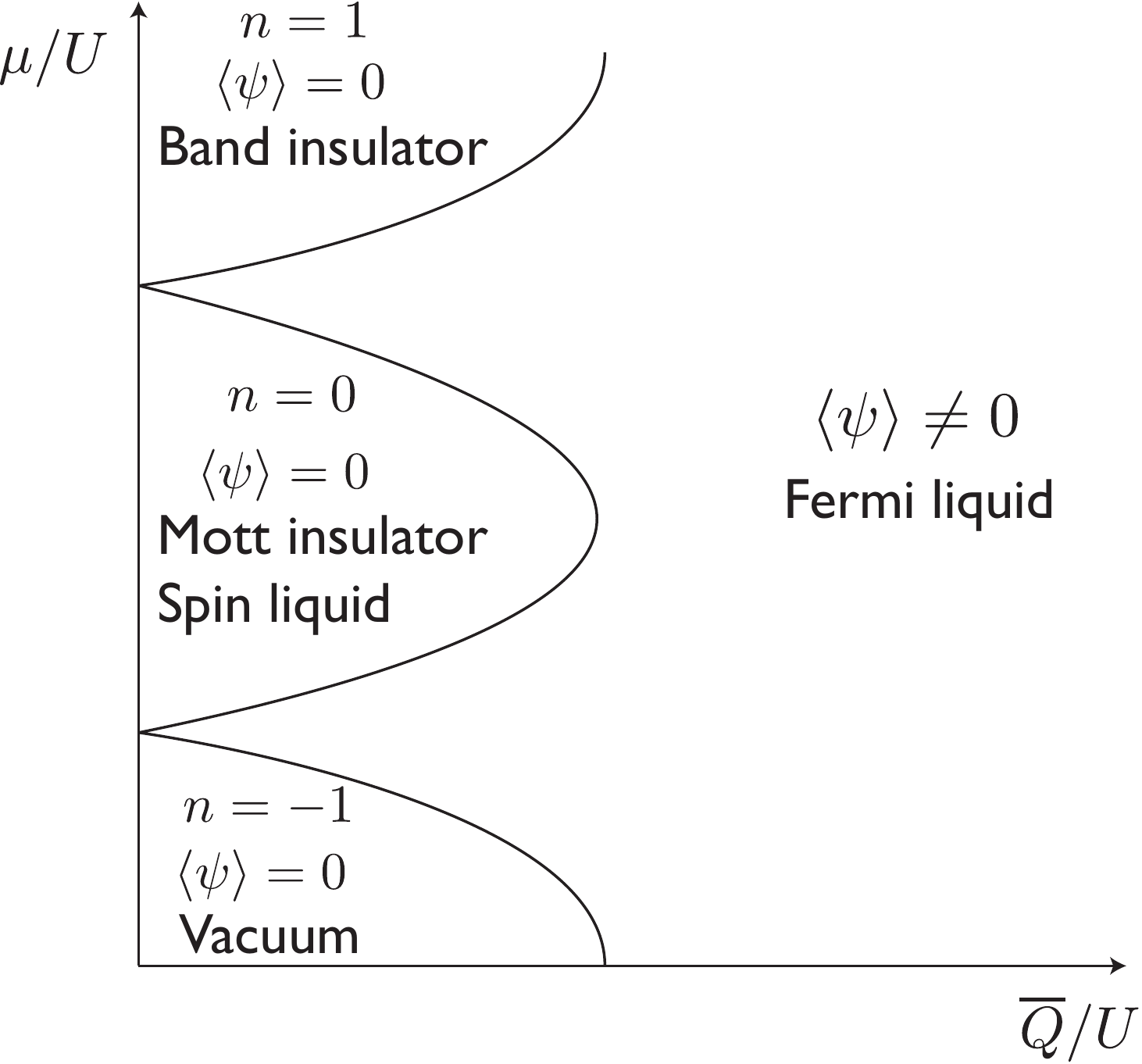}
 \caption{Possible phase diagram of the electron Hubbard model in Eq.~(\ref{h1}) on the triangular lattice.
 This phase diagram is obtained by a mean-field analysis  of the theory $\mathcal{L}_b$ in Eq.~(\ref{rotorfermiongauge}), 
 similar to that for the boson Hubbard model in Chapter 9 of Ref.~\onlinecite{ssbook2p}.
 We use the number operator $n = p^\dagger p - h^\dagger h$, which commutes with the boson hopping term,
 to characterize the Mott insulating states.
 Only the Mott insulating lobes with $n = -1,0,1$ are compatible with the constraint in Eq.~(\ref{fnc}); these Mott insulating
 lobes have fermion density $\left\langle f_\alpha^\dagger f_\alpha \right\rangle = n + 1$. }
\label{bosemott}
\end{figure}

The phase diagram of $\mathcal{L}_b$ is more interesting: it involves strong interactions between the $p$
and $h$ bosons.
It can be a analyzed in a manner similar to that of the boson Hubbard model (see Chapter 9 of Ref.~\onlinecite{ssbook2p}),
leading to the familiar ``Mott lobe'' structure shown in Fig.~\ref{bosemott}.

At large values of $\overline{Q}/U$ we have the analog of the superfluid states of the boson Hubbard model, in which there is a condensate of the same operator as that in Eq.~(\ref{eq:lc}):
\beq
\psi = p + h^\dagger.
\eeq
Note that $\psi$ is the ladder operator for the number operator $n= p^\dagger p - h^\dagger h$ 
used to characterize the insulating phases in Fig.~(\ref{bosemott}). 
The $\psi$ operator carries unit charge under the U(1) gauge field (from Eq.~(\ref{gauget}), 
and so the superfluid phase, with $\langle \psi \rangle \neq 0$, does
not break any global symmetries (unlike the boson model of Section~\ref{sec:conf}). 
Instead it is a ``Higgs'' phase. In the presence of 
the Higgs condensate, the operator relation in Eq.~(\ref{cf}) implies that $c_\alpha \sim f_\alpha$, and so the $f_\alpha$ fermions carry
the same quantum numbers as the physical electron. Consequently, the $f_\alpha$ Fermi surface is simply an electron Fermi surface.
Furthermore, the Higgs condensate quenches the $B_\mu$ fluctuations, and so there are no singular interactions between the 
Fermi surface excitations. This identifies the present phase as the familiar Fermi liquid, 
as noted in Fig.~\ref{bosemott}. We note that we can equally well identify this phase as a ``confining'' phase of the U(1)
gauge theory, in which the $\psi$ boson has formed a bound state with the $f_\alpha$ fermions, which is just the
gauge-neutral $c_\alpha$ fermion. Indeed, as is well known, Higgs and confining phases are qualitatively the same when
the Higgs condensate carries a fundamental gauge charge, as is the case here.

Having reproduced a previously known phase of the Hubbard model in the U(1) gauge theory, let us now examine the new phases
within the `Mott lobes' of Fig.~\ref{bosemott}. In these states, the boson excitations are gapped, and 
number operator $n = p^\dagger p - h^\dagger h$
has integer expectation values. The constraint in Eq.~(\ref{fnc}) implies that only $n = -1,0,1$ are acceptable values,
and so only these values are shown. It is clear from the representation in Eq.~(\ref{rotorrep}) that any excitation involving change in electron
number must involve a $p$ or $h$ excitation, and so the gap to the latter excitations 
implies a gap in excitations carrying non-zero electron number. This identifies
the present phases as insulators. Thus the phase boundary out of the lobes in Fig.~\ref{bosemott} is a metal-insulator transition.

The three insulators in Fig.~\ref{bosemott} have very different physical characteristics. 

Using the constraint in Eq.~(\ref{fnc}) we see that
the $n =-1$ insulator has no $f_\alpha$ fermions. Consequently this is just the trivial empty state of the Hubbard model,
with no electrons.

Similarly, we see that the $n=1$ insulator 
has 2 $f_\alpha$ fermions on each site. This is the just the fully-filled state of the Hubbard
model, with all electronic states occupied. It is a band insulator.

Finally, we turn to the most interesting insulator with $n =0$. 
Now the electronic states are half-filled, with $\langle f_\alpha^\dagger f_\alpha \rangle = 1$. Thus there is an unpaired fermion on each site, and its spin is free to fluctuate. There is a non-trivial wavefunction in the spin sector,
realizing an insulator which is a `spin liquid'. In our present mean field theory, the spin wavefunction is specified by Fermi surface state
of the $f_\alpha$ fermions. Going beyond mean-field theory, we have to consider the fluctuations of the $B_\mu$ gauge field,
and determine if they destabilize the spin liquid. 
The $f_\alpha$ fermions carry the $B_\mu$ gauge charge, and these
fermions form a Fermi surface.  The gapless fermionic excitations 
at the Fermi surface prevent the proliferation of monopoles in the compact U(1) gauge field: 
the low energy fermions suppress
the tunneling event associated with global change in $B_\mu$ flux\cite{hermelep,leemp}.
Thus the emergent U(1) gauge field remains in a deconfined phase, and this spin liquid state is stable.
These gapless gauge excitations have strong interactions with the $f_\alpha$ fermions, and this
leads to strong critical damping of the fermions at the Fermi surface
which is described by a strongly-coupled field theory\cite{leenp,metnemp,mrossp}.
The effect of the gauge fluctuations is also often expressed in terms of an improved trial wavefunction for
the spin liquid \cite{motp}: we take the free fermion state of the $f_\alpha$ fermions, and apply a projection
operator which removes all components which violate the constraint in Eq.~(\ref{fnc}). This yields the 
`Gutzwiller projected' state
\beq
|\mbox{spin liquid} \rangle = \left( \prod_i \left[ \frac{1 - (-1)^{\sum_\alpha f_{i \alpha}^\dagger f_{i \alpha}}}{2} \right] \right)
\left( \prod_{{\bm k} < k_F} f_{\uparrow}^\dagger ({\bm k} ) f_{\downarrow}^\dagger ({\bm k} ) \right) | 0 \rangle,
\eeq
where the product over ${\bm k}$ is over all points inside the Fermi surface. 

Finally, we turn to an interesting quantum phase transition in Fig.~\ref{bosemott}. This is the transition between the spin liquid
and the Fermi liquid at total electron density $\mathcal{N}=1$, which occurs at the tip of the $n =0$ Mott lobe. 
From the boson sector, this looks like a Higgs transition, of the condensation of a complex scalar $\psi$ as in
Section~\ref{sec:conf}, but in the presence of a fluctuating U(1) gauge field. 
However, the fermionic sector is crucial in determining the nature of this transition. Indeed, in the absence of the Fermi surface, this transition would
not even exist beyond mean field theory: this is because the U(1) gauge field is compact, and the scalar carries unit charge, and so the
confining and Higgs phases of this gauge theory are smoothly connected. So we have to combine the Higgs theory of a complex scalar
with the gapless Fermi surface excitations. We can obtain the field theory for this metal-insulator
transition by applying the methods of Section~\ref{sec:conf} to Eq.~(\ref{rotorfermiongauge}).
The analog of the condition $\Delta_p = \Delta_h$ needed to obtain a density of one electron
per site is now $\overline{\lambda} + \mu = 0$. In this manner,
 we find the field theory \cite{ffl2p,senthilmottp}
\bea
\mathcal{L} &=& | ( \partial_\mu + i B_\mu) \psi|^2 + s |\psi|^2 + u |\psi|^4 
+ i B_\tau \mathcal{N} \nonumber \\
&+& f^\dagger_\alpha \left[ \frac{\partial}{\partial \tau} - \varepsilon_F - i B_\tau - \frac{1}{2m} 
({\bm \nabla} -  i {\bm A})^2  \right] f_{\alpha}, \label{fermihiggs}
\eea
where the energy $\varepsilon_F$ is to be adjusted to yield total fermion density $\mathcal{N} = 1$. 
The transition is accessed by tuning $s$, and we move from a spin liquid with $\langle \psi \rangle = 0$ for $s > s_c$, to a Fermi liquid
with $\langle \psi \rangle \neq 0$ for $s<s_c$. Within the the spin liquid phase, we can safely integrate out the gapped
$\psi$ quanta, and so obtain the theory in Eq.~(\ref{ssl}) of the main text.
The critical properties of the theory at $s=s_c$ have been studied \cite{senthilmottp,rkkp},
and an interesting result is obtained: the Fermi surface excitations damp the gauge bosons so that they become ineffective
in coupling to the critical $b$ fluctuations. Consequently, the gauge bosons can be ignored in the $\psi$ fluctuations,
and the transition is in the universality class of the 2+1 dimensional XY model.
In other words, quite unexpectedly,
the critical theory is the same as that of the superfluid-insulator transition of Section~\ref{sec:conf}.
There are additional gapless excitations associated with the gauge field and the Fermi surface,
but these are irrelevant for the values of certain critical exponents.

\end{document}